\documentclass[sn-aps]{sn-jnl}% American Physical Society (APS) Reference Style
\usepackage[utf8]{inputenc}
\usepackage[T1]{fontenc}
\usepackage{amsmath, amsthm, amssymb}
\usepackage{mathpazo}
\jyear{2022}%

\theoremstyle{thmstyleone}%
\newtheorem{theorem}{Theorem}%  meant for continuous numbers
\newtheorem{proposition}[theorem]{Proposition}% 

\theoremstyle{thmstyletwo}%

\theoremstyle{thmstylethree}%
\newtheorem{definition}{Definition}%

\newcommand{\sref}[1]{Sec. \ref{#1}}

\newcommand{\eref}[1]{Eq.\hspace{0.025in}(\ref{#1})}

\begin{document}

\title[Three types of Landauer's erasure principle: a microscopic view] {Three types of Landauer's erasure principle: A microscopic view}

\author[1] {\fnm{Xavier} \sur{Oriols}}\email{xavier.oriols@uab.es}
\equalcont{These authors contributed equally to this work.}
\author[2] {\fnm{Hrvoje} \sur{Nikoli\'c}}\email{hnikolic@irb.hr}
\equalcont{These authors contributed equally to this work.}
\affil[1]{\orgdiv{Departament  d'Enginyeria Electr\`{o}nica}, \orgname{Universitat Aut\`{o}noma de Barcelona},
 \orgaddress{\street{Edifici QC}, \city{Cerdanyola del Vall\`{e}s}, \postcode{08193}, \state{Barcelona}, \country{Spain}}}

\affil[2]{\orgdiv{Theoretical Physics Division}, \orgname{Rudjer Bo\v{s}kovi\'c Institute},
 \orgaddress{\street{Bijeni\v{c}ka cesta 54}, \postcode{10000}, \state{Zagreb}, \country{Croatia}}}



\abstract{An important step to incorporate information in the second law of thermodynamics was done by Landauer, showing that the erasure of information implies an increase in heat. Most attempts to justify Landauer's erasure principle are based on thermodynamic argumentations. Here, using just the time-reversibility of classical microscopic laws, we identify three types of the Landauer's erasure principle depending on the relation between the two final environments: the one linked to a logical input 1 and the other to the logical input 0. The strong type (which is the original Landauer's formulation) requires the final environments to be in thermal equilibrium. The intermediate type giving the entropy change of $k_B \ln 2$ occurs when the two final environments are identical macroscopic states. Finally, the weak Landauer's principle, providing information erasure with no entropy change, when the two final environments are macroscopically different. Even though the above results are formally valid for classical erasure gates, a discussion on their natural extension to quantum scenarios is presented. This paper strongly suggests that the original Landauer's principle (based on the assumption of thermalized environments) is fully reasonable for microelectronics, but it becomes less reasonable for future few-atoms devices working at THz frequencies. Thus, the weak and intermediate Landauer's principles, where the erasure of information is not necessarily linked to heat dissipation, are worth investigating.} 

%%================================%%
%% Sample for structured abstract %%
%%================================%%


\maketitle

\section{Introduction}
\label{intro}

\begin{figure}[ht]
\centering
\includegraphics[width=1\linewidth]{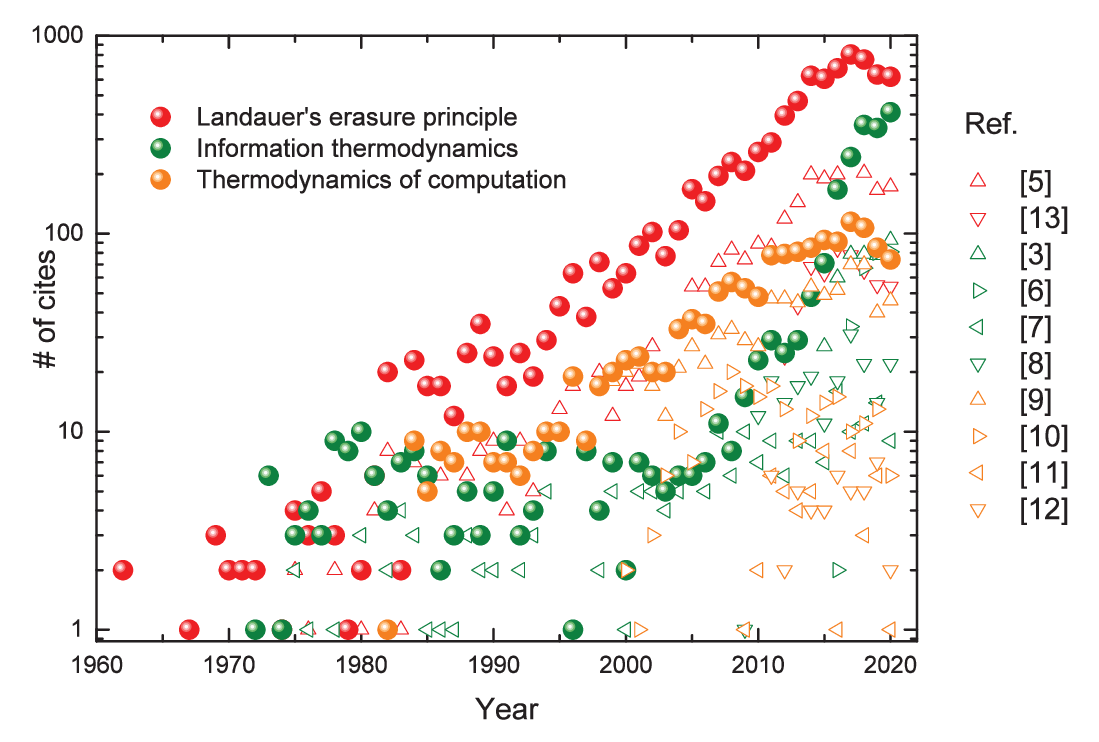}
\caption{Solid symbols denote the number of times cited as a function of year for the keywords \textit{Landauer's erasure principle} (red), \textit{Information thermodynamics} (green) and \textit{Thermodynamics of computation} (orange). Open symbols denote the same information for some of the relevant papers mentioned in the references of this manuscript. The seminal work of Landauer reached a maximum of attention in the literature when its first experimental validation by Berut et al. \cite{berut2012experimental}. The data are extracted from Ref. \cite{isi}.}
\label{References}
\end{figure}

For more than a century, important efforts have been devoted to understand the entropic and energetic costs of manipulating information. The first attempt for incorporating information into thermodynamics was as early as 1871 when James Clerk Maxwell presented the gedanken experiment, now known as Maxwell's demon \cite{maxwellbook} (a demon in the middle of a container with a trapdoor could transfer the fast and hot particles from a cold side to a hot one, in apparent violation of the second law of thermodynamics, if he had enough information about the particle velocities and positions). An analysis of Maxwell’s demon was conducted by Szilard \cite{szilard1929entropieverminderung} as early as 1929 when he studied an idealized heat engine with one particle gas and directly associated the information acquired by measurement with the physical entropy. Any practical implementation of the Maxwell's demon requires a finite memory to store information about decisions whether, for each particle, the trapdoor will be open or closed. Charles Bennett \cite{bennett1982thermodynamics}, and independently Oliver Penrose \cite{penrose1970}, clarified that the erasure of each bit of information in the memory requires a dissipation of heat in the environment, thus recovering the validity of the second law when the memory (demon) and its environment are properly included into the thermodynamic discussion. Bennett's and Penrose's conclusions were based on the previous work of Rolf Landauer \cite{landauer1961irreversibility} in 1961, showing that the erasure of information requires dissipation of a (minimum) amount of heat equal to $k_B T\ln 2$, where $k_B$ is the Boltzmann's constant and $T$ is the temperature. The work of Landauer is considered a key element on what Bennett named \textit{thermodynamics of computation} \cite{bennett1982thermodynamics,bennett1998informationC2,kahan2008towardsC3,qianefficientC4} or what nowadays is known by the more general term of \textit{information thermodynamics} \cite{parrondo2015thermodynamicsT1,vinjanampathy2016quantumT2,tribus1971energyT3,sagawa2009minimalT4} as seen in Fig.  \ref{References}.  

For the great majority of scientists, the seminal work of Landauer is a masterpiece of science \cite{bennett1973logical,berut2012experimental,bennett1982thermodynamics,bennett1988notes,bennett2003notes,lutz2015maxwells,frank2018physical,jacobs2005deriving,pezzutto2016implications,reeb2014improved,maruyama2009colloquium,bennett1984thermodynamically,benioff1984comment,toffoli1984commnet} connecting information and thermodynamics. Nevertheless, after more than 60 years, it is still accompanied by controversies. From a theoretical side, some scientists have persistently argued that the Landauer's principle is not a pertinent way of discussing dissipation in computing devices  \cite{norton2013all,norton2005eaters,norton2011waiting,norton2013end,hemmo2013entropy,kish2018information,porod1984dissipation,porod1984porod,kish2012energy}. 
From the experimental side, apart from recent successful experiments validating of the Landauer's erasure principle \cite{berut2012experimental,hong2016experimental,yan2018single,orlov2012experimental,berut2015information,berut2013detailed,ribezzi2019large}, there are few works suggesting some type of drawbacks \cite{gavrilov2016erasure,lopez2016sub}.

\begin{figure}[ht]
\centering
\includegraphics[width=1\linewidth]{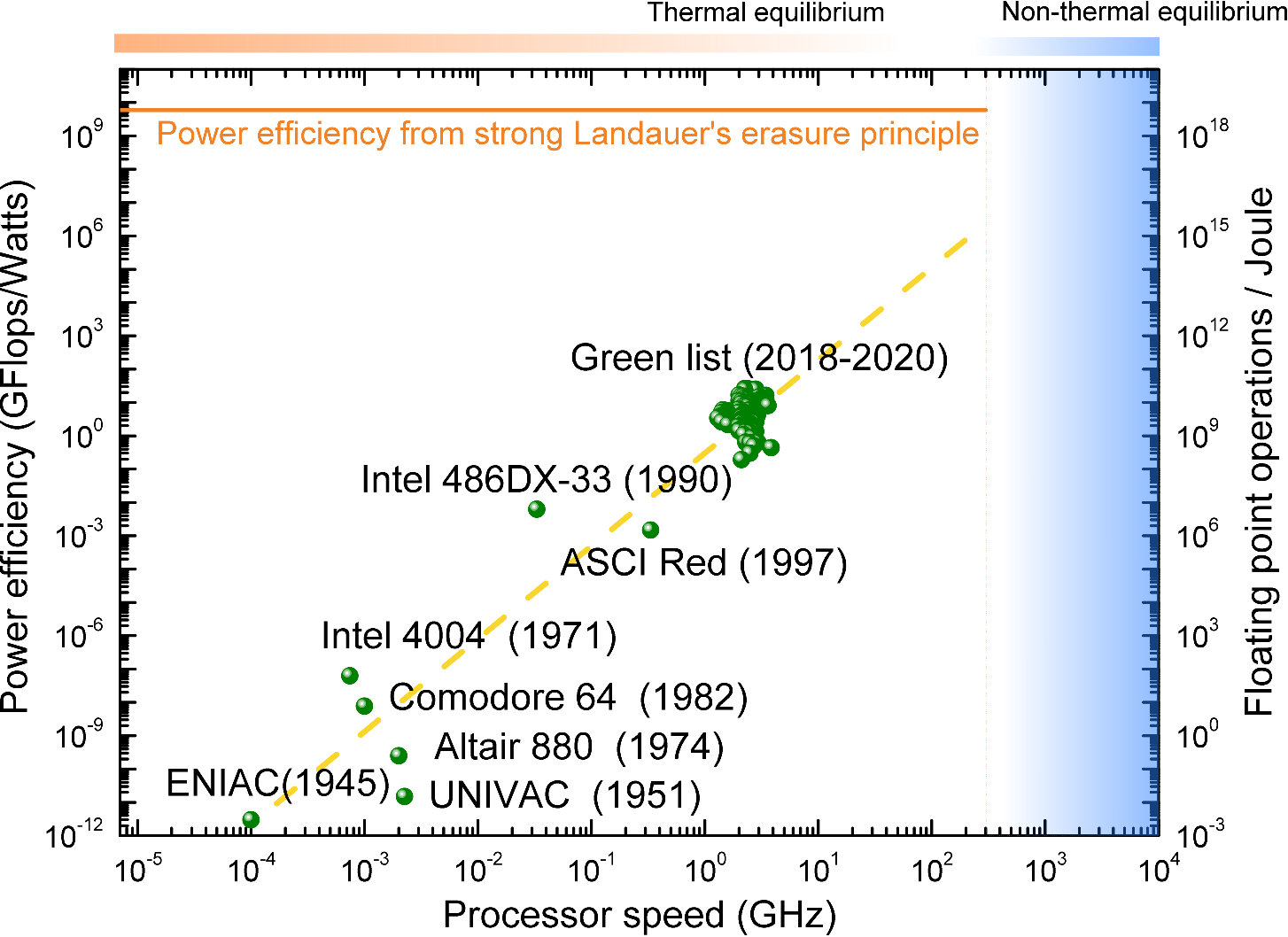}
\caption{Solid green symbols denote the power efficiency of different CPU's as a function of the processor speed. The orange line denotes the power efficiency limit of the strong (original) Landauer's erasure principle. The blue shaded region corresponds to processor speeds (operating frequencies) close or above 1 THz where the assumption of thermal environment is less evident. The yellow dashed line shows the tendency in last decades indicating that computers will reach the non-thermal equilibrium before reaching the Landauer limit, so that predictions of computing efficiency based on non-thermal reservoirs will become more relevant than the strong (original) Landauer's erasure limit. }
\label{historicaldissipation}
\end{figure}

The original motivation of Landauer's work as a part of his job as a researcher at the International Business Machines Corporation (IBM), however,  was not devoted to establish a link between information and thermodynamics, but just to find the minimal (if any) amount of heat dissipated by an \textit{ideal} computer. He brilliantly anticipated the minimum dissipation of $k_B T \ln 2$ per bit. The heat dissipated in computers is nowadays much larger, and it is the real bottleneck that prevents further progress. The power dissipated in electron devices is directly proportional to the working frequency. Thus, the higher frequency at which we make computations, the more heat is dissipated. The overall amount of power that can be dissipated from the chip imposes a limit on the operating frequency around 5 GHz for real computers, as seen in Fig. \ref{historicaldissipation}. In other words, the technology to build a single transistor working at frequencies as high as 1 THz is well developed, but not the technology to extract the amount of heat generated in a chip with $10^{10}$ of such transistors \cite{enrique2020charge}. In Fig. \ref{historicaldissipation} we indicate the power dissipation for the first computer built in 1945. It was named Electronic Numerical Integrator And Computer (ENIAC) \cite{eniac} and it required 174 kilowatts of power to run 5000 simple addition or 300 multiplications per second, with a clock rate of 100 kHz. The typical measure of computer performance is given in floating point operations per second (FLOPS). Although the ENIAC did not work with bits, we can estimate its computer performance around 500 Flops with a power efficiency of $3\cdot10^{-12}$ gigaflops/watt (see lower green dot Fig. \ref{historicaldissipation}). We compare ENIAC with nowadays supercomputers. In the November 2020 ranking of supercomputers in terms of energy efficiency \cite{Greenlist}, the NVIDIA DGX SuperPOD was the most energy-efficient supercomputer with 26.2 Gigaflops/watt. This demonstrates an awesome improvement of $12$ orders of magnitude in energy efficiency during last 75 years.  We can compare such numbers with Landauer's prediction by noticing that a floating-point operation reads in two numbers and returns one. If this is done on a computer with finite memory capacity, eventually the number which is being returned must erase another number in memory. Thus, according to Landauer's erasure principle stating that a dissipation of $k_B T\ln 2 \approx 2.8\cdot10^{-22}$ Joules is required by bit erased, one Joule of energy for an ideal Landauer computer would enable to re-write  $3.6\cdot10^{20}$  bits. Using 64 bits for a floating-point number, one Joule of energy would allow about  $6\cdot10^{18}$  floating point operations, which means $6\cdot10^{9}$ Gigaflops/watt. See the frequency-independent result in the orange line in Fig. \ref{historicaldissipation}. Certainly, the fact that even the most energy-efficient computers today are still $8$ orders of magnitude below the Landauer limit implies that the electronic industry needs to solve many problems before the Landauer's erasure principle becomes a relevant issue \cite{footnote8}. 

The overall message of the Landauer's erasure principle is that, even after developing the best technology in the future that will minimize the problems of heat dissipation in computers by $8$ orders of magnitude, we will still be faced with the fact that some heat dissipated ($k_B T \ln 2$ per bit) will not be an unnecessary nuisance, but a fundamental part of data erasure that cannot be avoided in any way, independently of the details of the computing device.   

The central topic of our paper is \textit{how fundamental} is the Landauer's erasure principle and if some type of extension (or generalization) is possible. In general, the Landauer's erasure principle is presented (and understood) in the literature as a fundamental result that cannot be avoided in any way. But, is it \textit{universally} true that, independently of the details of the computing device, a heat dissipation of $k_B T \ln 2$ per bit cannot be avoided when data are erased? We anticipate that the fact that the Landauer limit in Fig. \ref{historicaldissipation} is independent of the frequency (processor speed) is suspicious because the process of thermalization (the change from a non-equilibrium to an equilibrium thermodynamic state) is a dynamical process that requires some time in either classical or quantum reservoirs.

Since the Landauer's erasure principle is based on a thermodynamic explanation of computations, at first sight, it seems that the preliminary question that we have to answer is: \textit{how fundamental is thermodynamics?} Thermodynamics is a scientific discipline that explains complex systems through macroscopic properties, avoiding a need to discuss microscopic details. Historically, the thermodynamic laws were developed only for systems in the so-called thermodynamic equilibrium. In recent years, however, thermodynamics as a scientific theory has evolved to systems outside of thermodynamic equilibrium \cite{bookdavid}. The so called classical irreversible thermodynamics, under the hypothesis of local equilibrium, borrows most of the concepts and tools of equilibrium thermodynamics to non-equilibrium systems. Nowadays, even systems outside of local equilibrium are being studied in different branches of thermodynamics \cite{bookdavid}. Thus, whether a computing device represents a system that can be studied with some branch of thermodynamic is not a question. By the own flexibility of thermodynamics as a scientific discipline, it is always possible to construct a branch of thermodynamics with the ability to predict the macroscopic behavior of computing devices, even outside of thermodynamic equilibrium. Thermodynamics is becoming a \textit{science of everything} \cite{everything}, including a science of information thermodynamics.

The path followed in this paper to understand the universality (or the lack thereof) of the Landauer's erasure principle is a study of the erasure of information from a microscopic (mechanical) point of view, just by assuming the time-reversibility of microscopic laws, and then checking whether our general results (independent of any thermodynamic concepts) coincide or not with the original  Landauer's erasure principle. We show that depending on the type of final environment involved in the erasure of a logical $\textbf{1}$ or a logical $\textbf{0}$, three results can be established. The original Landauer's erasure principle, which we refer to as a {\em strong} type of Landauer's erasure principle, is recovered when the final state of the environment is in a thermodynamic equilibrium. Alternatively, an {\em intermediate} relation between manipulation of information and entropy change can be deduced when the only (macroscopic) condition imposed on the final environment is that they  \textit{look} indistinguishable (from a macroscopic point of view) when different logical inputs are involved. Such an intermediate relation gives the well-known limit  $k_B \ln 2$ of entropy change  when applied to an erasure gate. Finally, for states of environment that \textit{look} distinguishable we establish the {\em weak} type of Landauer's erasure principle which imply no entropy change for erasure computations. 

Thus, we conclude that the original (strong) Landauer's erasure result is not universal because thermal reservoirs are not universal. As we shall discuss in the last part of this paper, there are modern reservoirs/environments that never thermalize. Moreover, in case of thermalization, the dynamical transition from non-thermal to thermal reservoirs requires some time. In other words, as depicted by the shaded region of Fig. \ref{historicaldissipation}, the thermal reservoir assumption of the strong Landauer's erasure principle cannot be accepted uncritically for computing devices that switch from one state to the other faster than the time required to thermalize the reservoir. In the modern language of open systems \cite{opensystem}, these fast changing gate involve non-Markovian environments. As previously indicated, it is important to clarify that the limitations of the Landauer's erasure principle are not limitations of the \textit{information thermodynamics} itself because it is always possible to include some type of macroscopic effects of such non-Markovianity in thermodynamic formulations of computation, beyond the original Landauer's erasure principle.  

The structure of the rest of the paper is as follows. In Sec. \ref{definition} we define microscopic and macroscopic states, the physical characteristics of a logical gate and the requirements imposed by the time-reversibility of microscopic laws (Liouville theorem). In Sec. \ref{three} we define three types of the relation between manipulation of information and entropy change: strong, weak and an intermediate one, corresponding to three different types of final environments. Finally, we provide a discussion on how the above results can be extended to quantum systems in Sec. \ref{quantum}. We conclude in Sec. \ref{conclusions}. We also add two appendixes with technical details.    

\section{Definitions}
\label{definition}

In this section we provide detailed definitions of microscopic and macroscopic states in a general classical erasure gate.  The proper understanding of when a set of microscopic states is (or is not) identical to a macroscopic state will be the key-element in the developments of Sec. \ref{three}.

\subsection{Defining microscopic states}

We consider a closed (or isolated in the thermodynamic language) system with $N$ degrees of freedom. We distinguish the $N_S$ degrees of freedom of the system (the active region of the computing gate) and the $N_E=N-N_S$ environment degrees which represent all the \textit{other} degrees of freedom. The degrees of freedom of the system are represented by the vector $x$ with $6N_S$ components corresponding to three position and three momenta of each particle in the physical space. Similarly, the degrees of freedom of the environment are represented by $y$ as a vector in the $6N_E$-dimensional phase space of the environment \cite{footnote1}. The interaction between all degrees of freedom is determined by the (time-independent) Hamiltonian $H(x,y)$, which fully describes the physical implementation of the logical gate.

\begin{definition}[Microscopic state] 
 We define a microscopic state of the gate and environment at time $t$ by the point $x^{(j)}(t),y^{(j)}(t)$ in the $6N$-dimensional phase space $\Gamma$, where the superscript $j$ labels different solutions (corresponding to different experiments) from the same Hamiltonian $H(x,y)$. 
\end{definition}
In general, we will consider $j=1,..,M$ with $M$ large enough (but not infinite) so that the set of $x^{(j)}(t),y^{(j)}(t)$ is statistically meaningful. 

\subsection{Defining macroscopic properties and macroscopic states}

After the definition of microscopic states, we define here macroscopic properties and macroscopic states. 

\begin{definition} [Macroscopic property]
We define a macroscopic property as a function $\textbf{A}: \Gamma \to \mathbb{R}$ that assigns a real value to each point in the phase space $\Gamma$. Two phase-space points $x^{(j)}(t),y^{(j)}(t)$ and $x^{(k)}(t),y^{(k)}(t)$ are macroscopically identical (according to this property $\textbf{A}$) if and only if $\textbf{A}(x^{(j)}(t),y^{(j)}(t))=\textbf{A}(x^{(k)}(t),y^{(k)}(t))$.
\end{definition} 
Notice that there are no anthropomorphic implications in the definition of a macroscopic property \cite{footnote2}.  No human observation is needed. In our case, $\textbf{A}$ can be a the logical information of the system denoted by the logical symbols \textbf{0} and \textbf{1}. One can define these macroscopic properties as a result of a large-scale resolution of the apparatus involved in the identification of such property $\textbf{A}$. There is a large set of microscopic states at the output of the gate that are correctly interpreted as belonging to the logical \textbf{0} in the input of another subsequent gate. For a simple and objective definition, for example, a maximum distance from a central phase-space point can be used to specify which microscopic states belong to a given macroscopic property.  

Once we have a defined macroscopic property, we can define a macroscopic state. 

\begin{definition} [Macrostate] 
We define the macroscopic state (or macrostate) $A$ at time $t$ as the set of all microscopic states $x^{(j)}(t),y^{(j)}(t)$ that have the same macroscopic property $\textbf{A}$ at that time, namely
\begin{eqnarray}
A=\{\text{All } x^{(j)}(t),y^{(j)}(t) \in \Gamma  \text{ so that } \textbf{A}(x^{(j)}(t),y^{(j)}(t))=\textbf{A}\} .\nonumber
\end{eqnarray}
\label{macro}
Notice that $A$ is a subspace of $\Gamma$, while (bold) $\textbf{A}$ is just a number in real space. 
\end{definition}

We are now interested in defining the phase-space volume $V_A$ of the macrostate $A$. 

\begin{definition} [volume of macrostate $A$]
We define the volume of the macrostate $A$ by counting the number of microscopic states that satisfy the condition $\textbf{A}(x^{(j)}(t),y^{(j)}(t))=\textbf{A}$ in definition \ref{macro} as
\begin{eqnarray}
V_A= \Delta \Gamma \times M_A  \;\;\;\;\;\;\text{with}\;\;\;\; M_A=\sum_{j=1}^M \delta_{\textbf{A}(x^{(j)}(t),y^{(j)}(t)),\textbf{A}} ,\nonumber
\end{eqnarray}
where $\delta_{a,b}$ is the Kronecker delta function (that becomes one when $\textbf{A}(x^{(j)}(t),y^{(j)}(t))=\textbf{A}$) and $\Delta \Gamma$ is an irrelevant phase-space volume small enough to accommodate zero or one microstate (see appendix A). We remind that $M$ is large enough (but not infinite) so that the results are statistically meaningful. 
\label{va}
\end{definition} 

We will also be interested in identifying those degrees of freedom of the system alone that belong to the set $A$. We define the system subspace $X_A$ as the set of microscopic points in the system phase space $\Gamma_S$ that belong to $A$, as $X_A=\{\text{ All } x^{(j)}(t) \in \Gamma_S  \text{ so that } x^{(j)}(t),y^{(j)}(t) \in A \}$.  Similarly, for the points in the environment phase space $\Gamma_E$, we define the Environment subspace as $Y_A=\{\text{ All } y^{(j)}(t) \in \Gamma_E  \text{ so that } x^{(j)}(t),y^{(j)}(t) \in A \}$ where the whole phase space is just the product of the system and environment phase spaces, $\Gamma=\Gamma_S \times\Gamma_E$.

\subsection{Physical gate as a transition between microscopic states} 

For each $j$-experiment, the Hamiltonian $H(x,y)$ determines the trajectory in the 6N-dimensional phase space between the initial values $x^{(j)}(t_i),y^{(j)}(t_i)$ and the final values $x^{(j)}(t_f),y^{(j)}(t_f)$. 

\begin{definition} [operation]
We define an operation or evolution of the states due to the Hamiltonian $H(x,y)$ as a bijective (one-to-one and onto) map $h(t_f,t_i)$ from the phase space $\Gamma$ at time $t_i$ (domain) to the same phase space $\Gamma$ at time $t_f$ (range) 
\begin{eqnarray}
h(t_f,t_i): A \to B.\nonumber
\end{eqnarray}
\label{bij}
where $B$, as the image of $A$ under the bijective mapping $h(t_f,t_i)$, is defined as
\begin{eqnarray}
B=\{ \text{All } x^{(j)}(t_f),y^{(j)}(t_f) \in \Gamma  \text{ so that } x^{(j)}(t_i),y^{(j)}(t_i) \in A\} \nonumber
\end{eqnarray}
\end{definition}
We notice that no macroscopic property $\textbf{B}$ is used in the description of $B$ as image of $A$ in the definition \ref{bij}. In other words, the set of microscopic states at the initial time $t_i$, that define a macroscopic state $A$, does not need to be a macroscopic state of the same macroscopic property  $\textbf{A}$ at the latter time $t_f$. 

\begin{proposition} The phase-space volume $v_B(t_f)$ of $B$, defined as image of $A$, satisfies  $v_B(t_f) \equiv V_A(t_i)$. 

The proof is simple. By construction, the states that belong to $B$ at $t_f$ are just the states that belonged to $A$ at time $t_f$. 
\begin{eqnarray}
v_B(t_f) \equiv V_A(t_i) \equiv \Delta \Gamma \times M_A  \;\;\;\;\;\;\text{with}\;\;\;\; M_A=\sum_{j=1}^M \delta_{\textbf{A}(x^{(j)}(t_i),y^{(j)}(t_i)),\textbf{A}}\nonumber.
\end{eqnarray}
This is, in fact, a simpler way of stating the Liouville theorem \cite{goldstein2018mecanica}. 
\label{vb}
\end{proposition}

We insist that the (non-capital) volume $v_B(t_f)$ do not need to be the volume of a macroscopic state $A$ at the final time $t_f$ defined as $V_A(t_f)$. It is possible that  $V_A(t_f) \ne V_A(t_i)$ if the microscopic states that satisfy $\textbf{A}(x^{(j)}(t_i),y^{(j)}(t_i))=\textbf{A}$ at the initial time $t_i$ are not the same states that satisfy the condition $\textbf{A}(x^{(j)}(t_f),y^{(j)}(t_f))=\textbf{A}$ at the final time $t_f$.  

Obviously, such evolution of microscopic states encodes an evolution of the logical information as well.  

\begin{definition} [Physical gate]
We define a gate at the physical level (with one bit of information that can take two initial logical values) as the following two maps: 
\begin{itemize}  
\item A map $h_{\textbf{0}}(t_f,t_i)$ when the involved initial microscopic states $A$ are those belonging to the information $\textbf{0}$
\item A map $h_{\textbf{1}}(t_f,t_i)$ when the involved initial microscopic states $A'$ are those belonging to the information $\textbf{1}$. 
\end{itemize}
By construction, such a composed map is also a bijective (one-to-one and onto) map from $\Gamma \times \Gamma$ to $\Gamma \times \Gamma$
\begin{eqnarray}
h_{\textbf{0}}(t_f,t_i) \times h_{\textbf{1}}(t_f,t_i): 
A \times A'  \to B \times B' .\nonumber
\end{eqnarray}
\label{bij2}
In fact, the bijective maps $h_{\textbf{0}}(t_f,t_i)$ or $h_{\textbf{1}}(t_f,t_i)$ mean, at the physical level, that microscopic classical laws are time-reversible \cite{goldstein2018mecanica}. If two phase-space trajectories coincide at one time, then such trajectories are identical at all times. This time-reversibility has important consequences on the type of physical transitions that are allowed. 
\end{definition}

We are now in conditions to present the following proposition that will be important along the paper: 
\begin{proposition}
If two different operations, $A \to B$ and $A' \to B'$ (where $B$ and $B'$ are the images of $A$ and $A'$, respectively), have a null intersection at the initial time, then they have a null intersection at any time.  In other words,
\begin{eqnarray}
\text{if}\;\; A  \cap A' =\emptyset \;\;\; \text{then} \;\;\;\; B  \cap  B' =\emptyset. \nonumber
\end{eqnarray}
\label{empty}
The demonstration is simple. Let us imagine that $ B  \cap  B'  \neq \emptyset$ because $x^{(j)}(t_f),y^{(j)}(t_f)=x^{(k)}(t_f),y^{(k)}(t_f)$. Then, because of time-reversibility, such trajectories are identical at the initial time too, i.e. $x^{(j)}(t_i),y^{(j)}(t_i)=x^{(k)}(t_i),y^{(k)}(t_i)$, so that  $A  \cap A' \neq \emptyset$ too. 
\end{proposition}

Notice that we have used in the proposition \ref{empty} the fact that different trajectories, for example $x^{(j)}(t_f),y^{(j)}(t_f)$ and $x^{(k)}(t_f),y^{(k)}(t_f)$, do not cross in phase-space at any time. This will be the condition that we will check in any proposal of a gate. Notice that physical systems defined from the Hamiltonian $H(x,y)$ do always satisfy this proposition \ref{empty}. But, the proposition \ref{empty} is also true for any (non-Hamiltonian) dynamical system that preserves phase-space volumes.  As a consequence, the three Landauer's erasure principles presented in this paper can be relevant, not only for the physical gates linked to $H(x,y)$ studied in this paper, but for applications in other areas outside physics described from divergenceless models.   

\subsection{Logical gate as transition between macroscopic states} 

From the logical information alone (forgetting about the microscopic state), we define the gate from a logical point of view as:
\begin{definition} [Logical gate]
We define a logical gate as a map $i(t_f,t_i)$ from the logical space $\mathbb{L}=\{\textbf{0},\textbf{1}\}$ at time $t_i$ (domain) to the same logical space $\mathbb{L}$ at time $t_f$ (range) as
\begin{eqnarray}
i(t_f,t_i): \textbf{A},\textbf{A'}  \to \textbf{C},\textbf{C'} . \nonumber
\end{eqnarray}
\label{nobij}
The logical information (or macroscopic property) \textbf{A} or \textbf{A'} can be perfectly identified with the macroscopic state $A$ or $A'$ in the definition \ref{bij2} of  a physical gate. However,  the logical information (or macroscopic property) \textbf{C} or \textbf{C'} in this new definition of a logical gate cannot be identified with the image states $B$ or $B'$ as defined in \ref{bij2}. 
\end{definition}
The difference between the physical and logical gate, which is the central point in our future discussion, can be translated to saying that, contrary to the bijective mapping $h(t_f,t_i)$ for microscopic states, the new logical map for macroscopic states $i(t_f,t_i)$ is not bijective. For example, we will be interested in two operations that define an erasure gate: the $\textbf{0}\to \textbf{0}$ operation and the $\textbf{1}\to \textbf{0}$ operation. Clearly, the map is not bijective. In the language of computation, it is said that the erasure gate is the simplest example of logical irreversibility, because the final (logical) information $\textbf{0}$ does not allow us to deduce what was the initial (logical) information (either $\textbf{0}$ or $\textbf{1}$). 

We want to clarify why we said in definition \ref{bij2} that $A,A'  \rightarrow B,B'$ is physically reversible (bijective), while we said in definition \ref{nobij} that $\textbf{A},\textbf{A'}  \rightarrow \textbf{C},\textbf{C'}$ can be logically irreversible (not bijective). The first refers to the evolution of microscopic states and the second to the evolution of macroscopic states. By construction, it is possible to find two different phase-space points $\{x^{(j)}(t_i),y^{(j)}(t_i)\}\neq\{x^{(k)}(t_i),y^{(k)}(t_i)\}$ that have the same (logical) information $\textbf{A}(x^{(j)}(t_i),y^{(j)}(t_i))=\textbf{A}(x^{(k)}(t_i),y^{(k)}(t_i))$, but it is not possible to find two identical (or very similar) phase-space points  $\{x^{(j)}(t_i),y^{(j)}(t_i)\} \approx \{x^{(k)}(t_i),y^{(k)}(t_i)\}$ that have different (logical) information $\textbf{A}(x^{(j)}(t_i),y^{(j)}(t_i)) \neq \textbf{A}(x^{(k)}(t_i),y^{(k)}(t_i))$.

\section{Three types of Landauer's erasure principle}
\label{three}

 Next, we distinguish three types of the relation between the erasure of information and its energetic and entropic costs, corresponding to three types of relation between the two final environments: the final environment belonging to the logical operation  $\textbf{1}\to \textbf{0}$ and the one to $\textbf{0}\to \textbf{0}$. Only the third type is the one developed originally by Landauer (in terms of a thermal reservoir).  We still keep the name (intermediate and weak) Landauer's erasure principle for the other two because we believe that we follow the original motivation of Landauer: encoding information in macroscopic properties and analyzing how the distribution of microscopic states that build such macroscopic state change during the erasure procedure. But our approach differs from the original one in the sense that we assume nothing more than time-reversibility of the microscopic laws. 

We assume that the gate is characterized by the logical information $\textbf{0}$ or $\textbf{1}$ (or the corresponding macroscopic states $A$ and $A'$ in definition \ref{bij2}), while the environment is characterized by another macroscopic property $\textbf{E}_\textbf{0}$ or $\textbf{E}_\textbf{1}$ (or its corresponding macroscopic states $E$ in definition \ref{macro}). Since a gate involves two operations, whenever needed we will specify which operation we are referring to by using, for example in an erasure gate, the label $\textbf{1}\to \textbf{0}$ or $\textbf{0}\to \textbf{0}$. We will also specify the time at which we are defining the macroscopic properties or states, by writing $t_i$ for the initial time and $t_f$ for the final one. 

\subsection{The weak Landauer's erasure principle}
\label{weak}

We first consider erasure gates where the final environments are macroscopically different at the final time:

\begin{itemize}
\item \textbf{Condition C1}: ENVIRONMENTS WITH FINAL DIFFERENT MACROSCOPIC PROPERTIES.  For two
different operations involved in a gate with two initial environment states 
which have macroscopically identical properties at the initial time
(e.g. $\textbf{E}_{\textbf{1}\to \textbf{0}}(t_i)=\textbf{E}_{\textbf{0} \to \textbf{0}}(t_i)$), 
the two final environment states have different macroscopic properties $\textbf{E}_{\textbf{1}\to \textbf{0}}(t_f) \ne\textbf{E}_{\textbf{0} \to \textbf{0}}(t_f)$) at the final time \cite{footnote2,footnote2bis}.  
\end{itemize}

\begin{figure}[ht]
\centering
\includegraphics[width=0.6\linewidth]{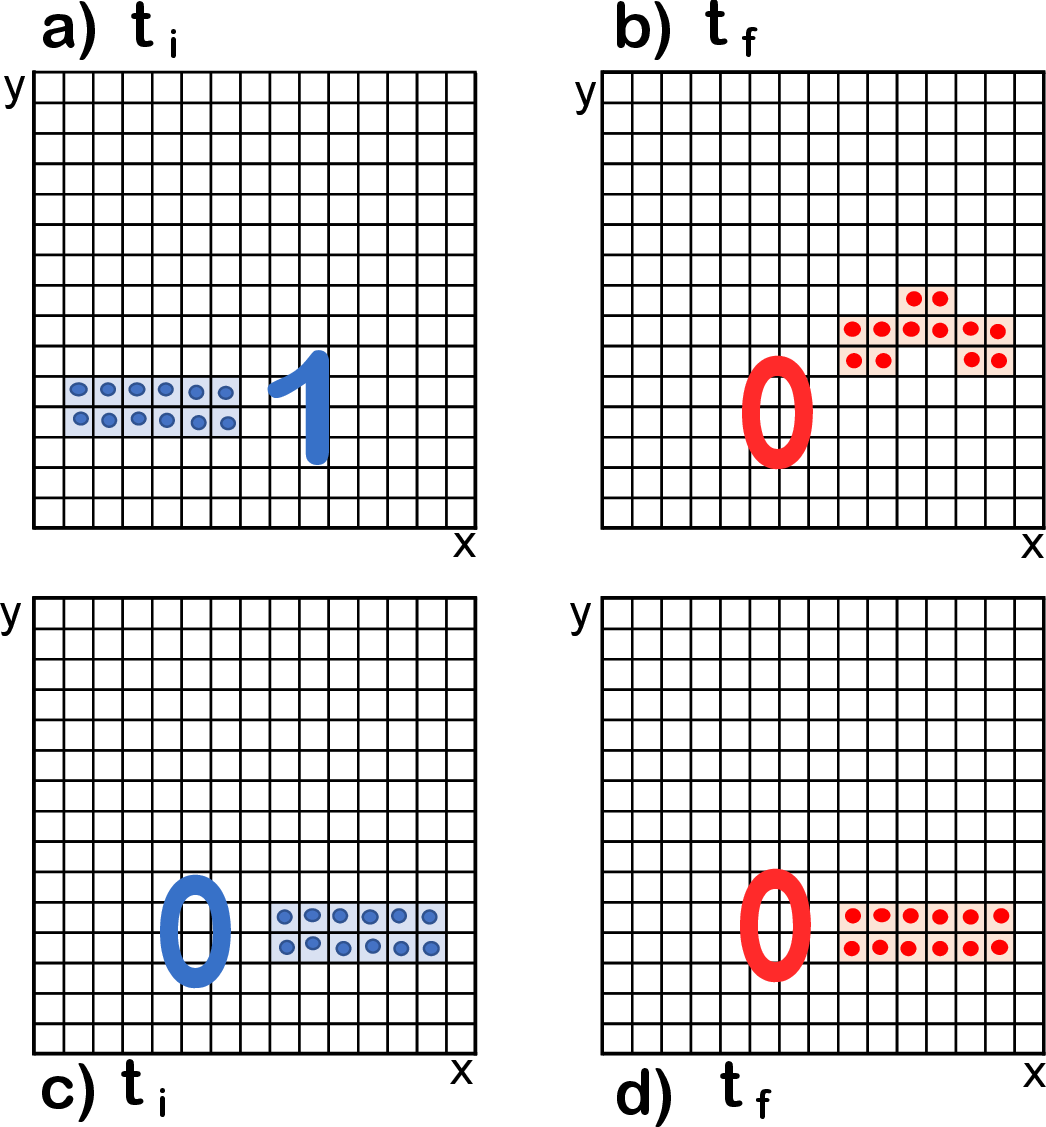}
\caption{Schematic representation of the initial (left panels) and final (right panels) microscopic states (dark blue and red solid circles) and the volumes of macrostates (light blue and red regions) in the system $x$ plus environment $y$ phase space. The upper panels correspond to the operation $\textbf{1}\to\textbf{0}$ and the lower panels to $\textbf{0}\to\textbf{0}$.  The macroscopic property of the system $\textbf{1}$ means being on the left of the $x$ axis, while the macroscopic property of the systems $\textbf{0}$ means being on the right of $x$ axis. The initial macroscopic environment properties (in the left panels) are identical $\textbf{E}_{\textbf{1}\to \textbf{0}}(t_i) = \textbf{E}_{\textbf{0} \to \textbf{0}}(t_i)$, while we can distinguish the macroscopic properties of the environment at the final time (in the right panels) $\textbf{E}_{\textbf{1}\to \textbf{0}}(t_f) \ne\textbf{E}_{\textbf{0} \to \textbf{0}}(t_f)$. Even though the gate is logically irreversible, it satisfies the time-reversibility of microscopic laws. The relevant point is that condition \textbf{C1} shows that each operation of the erasure process can be done without any change in the global entropy: the phase-space volumes (entropies) of the initial macroscopic states (in light blue regions in the left panels) are equal to the phase-space volumes (entropies) of the final macroscopic states (in the light red regions in the right panels). }
\label{schemeweak}
\end{figure}

Let us analyze \textbf{C1} for an erasure gate in Fig. \ref{schemeweak}. The initial logical states $\textbf{1}$ in Fig. \ref{schemeweak}(a)  and $\textbf{0}$ in Fig. \ref{schemeweak}(c) are different macroscopically (being in the left and in the right respectively), while having the same environment macroscopic properties and states, $E_{\textbf{1}\to\textbf{0}}(t_i)= E_{\textbf{0}\to\textbf{0}}(t_i)$ and $Y_{\textbf{0}\to \textbf{0}}(t_i)=Y_{\textbf{1}\to \textbf{0}}(t_i)$. By contrast, the final logical states $\textbf{0}$ in Fig. \ref{schemeweak}(b)  and $\textbf{0}$ in Fig. \ref{schemeweak}(d) are macroscopically identical (being both on the right), $X_{\textbf{0}\to \textbf{0}}(t_f)=X_{\textbf{1}\to \textbf{0}}(t_f)$, while having different environment macroscopic properties and states, $E_{\textbf{1}\to\textbf{0}}(t_f)\ne E_{\textbf{0}\to\textbf{0}}(t_f)$ and $Y_{\textbf{0}\to \textbf{0}}(t_f) \ne Y_{\textbf{1}\to \textbf{0}}(t_f)$. We clearly satisfy the proposition \ref{empty} at all times so that such an erasure process is possible from our mechanical point of view. Notice that condition \textbf{C1} implies that the initial macroscopic information will effectively disappear from the final state of the system, but it will appear in the final environment state. These results just show that, due to time-reversibility of microscopic laws, information can never be erased at the microscopic level in a full closed system. We note that we have arrived to the same conclusion as Hemmo and Shenker \cite{hemmo2013entropy}, but within a framework that will allow us to reach Landauer's and Bennett's results in a general and compact unified framework. 

\begin{proposition}
The erasure of information with a gate satisfying condition \textbf{C1} is compatible with no entropy cost $\Delta S=0$. 

For the proof, we use the Boltzmann entropy defined as the number of microstates that correspond to a macrostate (as discussed in the appendix A). We define $V_{\textbf{1}\to \textbf{0}}(t_i)$ as the phase-space volume of the  initial macrostate 
$X_{\textbf{1}\to \textbf{0}}(t_i), Y_{\textbf{1}\to \textbf{0}}(t_i)$. Similarly, we define  
$V_{\textbf{0}\to \textbf{0}}(t_i)$ as the phase-space volume of $X_{\textbf{0}\to \textbf{0}}(t_i), Y_{\textbf{0}\to \textbf{0}}(t_i)$. In the logical operation $\textbf{1}\to \textbf{0}$, the condition \textbf{C1} is compatible with defining the number of microstates of the final macrostate, $V_{\textbf{1}\to \textbf{0}}(t_f)$, equal to the number of microstate of the image of the initial macrostate $v_{B,\textbf{1}\to \textbf{0}}(t_f) \equiv V_{\textbf{1}\to \textbf{0}}(t_i) =V_{\textbf{1}\to \textbf{0}}(t_f)$. Identically, in the logical operation $\textbf{0}\to \textbf{0}$, we define $v_{B,\textbf{0}\to \textbf{0}}(t_f) \equiv V_{\textbf{0}\to \textbf{0}}(t_i) =V_{\textbf{0}\to \textbf{0}}(t_f)$. Then
\begin{eqnarray}
\Delta S&=&(p)\Delta S_{\textbf{1}\to \textbf{0}}+(1-p)\Delta S_{\textbf{0}\to \textbf{0}}\nonumber\\
&=&(p) k_B \left( \ln\big( V_{\textbf{1}\to \textbf{0}}(t_f)\big) - \ln \big(V_{\textbf{1}\to \textbf{0}}(t_i) \big)\right)\nonumber\\
&+&(1-p) k_B \left( \ln\big( V_{\textbf{0}\to \textbf{0}}(t_f)\big) - \ln\big( V_{\textbf{0}\to \textbf{0}}(t_i) \big)\right)=0. 
\end{eqnarray}
We have assumed an arbitrary probability $p$ for the $\textbf{1}\to\textbf{0}$ and $1-p$ for $\textbf{0}\to\textbf{0}$ operations. The reason why $\Delta S=0$ is possible is because the condition \textbf{C1} itself imposes that the environments are microscopically different, so that $v_{\textbf{1}\to \textbf{0}}(t_f) \cap v_{\textbf{0}\to \textbf{0}}(t_f)=\emptyset$ as required from  proposition \ref{empty}. Thus, the condition $\Delta S=0$ does not violate any fundamental microscopic law and a Hamiltonian $H_{\textbf{C1}}(x,y)$ is possible \cite{footnote3bis}.
\end{proposition}
The proof is just a consequence that the initial and final macrostates (seen as light blue and light red regions for the initial and final times,  respectively, in Fig.  \ref{schemeweak}) always have the same number of microstates. This fact is the well-known result given by the Liouville theorem \cite{goldstein2018mecanica} when dealing with $A$ and the image of $A$ at a later time. We remind the reader that, in more general scenarios, the number of microscopic points that are part of the macroscopic property $\textbf{A}(t_i)$ at the initial time does not need to be equal to the number of points that are part of the macroscopic property $\textbf{A}(t_f)$ at the final time. 

In fact, reading carefully the original works of Landauer and Bennett, one notices that the possibility of such types of erasure gates, giving $\Delta S=0$, was already well known to Landauer and Bennett. Bennett mentioned what he thought was the problem with such types of erasure gates in his 1973 paper \cite{bennett1973logical}. He erroneously concluded that it was not possible to use such erasure gates with condition \textbf{C1} more than once, because the environment is different each time we use the erasure gate (see the macroscopic states of environment in Fig.  \ref{schemeweak} (b) and (d)). Contrary to the Bennett's conclusion, we argue here that such erasure gate with $\Delta S=0$ can be used as many times as typical erasure gates can.  It is erroneously argued in \cite{bennett1973logical} that such an erasure gate will require a reset of the environment to its initial state to make the erasure gate useful again. However, we notice that in a conventional gate, in fact, the initial macroscopic state of environment is not identical to the final macroscopic state of environment: the final one contains more heat than the initial one. And yet, no reset to the initial cooler environment is assumed each time the gate is used. Similarly, we can assume that the change of state of the environment in the gate of Fig.  \ref{schemeweak} is small enough to be used again without reset \cite{footnote6}. See also appendix \ref{appb} with a toy model of an erasure gate satisfying  condition \textbf{C1}. This toy-model gate works properly without reset more than 30 times despite the fact that the environment is modified each time in such a way that one can guess what was the initial logical value by just looking at the environment variation. 

The true reason why the erasure gate depicted in Fig \ref{schemeweak} and conventional gates can be used many times is because of the change in the environment degrees of freedom $y$.  In other words, it is mandatory to change the microscopic degrees of freedom of the environment each time an erasure process takes place.  Because of the time reversibility of Hamiltonian dynamics, two initially different trajectories of the system alone without environment, $x^{(j)}(t_i)={\bf 1}$ and $x^{(k)}(t_i)={\bf 0}$, cannot become identical later, $x^{(j)}(t_i)=x^{(k)}(t_f)={\bf 0}$. A way to use such erasure gates (with $\Delta S=0$ or with $\Delta S \ne 0$) is to require $y(t)$ to be different each time we use the gate, but not too different. Finally, notice that the role played by the \textit{environment} $y(t)$ in the gates under condition \textbf{C1} are quite similar to the role played by the \textit{control register} in the gates of the \textit{reversible computation} proposed by Bennett \cite{bennett1984thermodynamically,bennett2003notes,bennett1988notes}. In both cases, the \textit{environment} or the \textit{control register} is the additional degree of freedom $y(t)$ needed to erase the system information $x(t)$ without violating the time-reversibility of the whole system. The difference is that the \textit{control register} is an active element in \textit{reversible computation}, while the \textit{environment} is interpreted here as a passive element without requiring any attention (reset). 

Fig \ref{schemeweak} shows an example on how to realize irreversible logic with reversible physics. We are requiring the final environments to be \emph{slightly} different at the macroscopic level. See also appendix \ref{appb} with a toy model of an erasure gate that works properly without reset. Certainly, our simplified erasure gate has a limit on the number of times it can be used. But, in principle, it is not different from conventional erasure gates in our computers, because they also have a limit on the number of times that they can be consecutively used, which is related to the limit on the extra heat that can be absorbed by the environment when we take into account that the number $N_E$ of environment particles is not strictly infinite. 

Does the result obtained above, where the information is erased without entropy cost, violate the original Landauer's prediction? Is the \textit{exorcise} of the Maxwell demon done by Bennett and Penrose, based on prior Landauer's cost for erasing data, wrong? We notice that our environments in \textbf{C1}, as plotted in Fig. \ref{schemeweak},  are not in thermodynamic equilibrium, so our results, as such, are not pertinent to discussions about systems that have assumed the hypothesis of thermodynamic equilibrium. Arguments on why we can expect non-thermal environment in some experiments will be discussed in more detail in \sref{quantum}.
However, as discussed in the introduction, thermodynamics is a scientific discipline flexible enough to accommodate these new results into a new (irreversible or non-equilibrium) branch of thermodynamics. 

Finally, the reader can argue that a fair discussion of the environment in present-day real computers has to involve a really large number of degrees of freedom ($N_E\gg 10^{23}$) making almost impossible to distinguish final environments, contrary to what we have stated in the \textbf{C1} condition and in appendix \ref{appb}. Sure, there are many environments in our ordinary life that can be considered as thermal environments. But, we will show in the last section of this paper that recent experiments in equilibration of closed quantum systems show environments that never thermalize or that the transition from a non-thermal to a thermal reservoir (for those which thermalize) needs some time. Thus, gates at very high frequency can imply (non-Markovian) environments that have not enough time to thermalize (to become independent of their initial conditions $\textbf{1}$ or $\textbf{0}$). There is a huge difference between saying that condition \textbf{C1} is  technologically difficult to reach,  and saying that \textbf{C1} is impossible to reach because it violates fundamental laws. In summary, there is no fundamental reason to expect that only thermal reservoirs can be applied to computations, so there is no reason to expect that the original Landauer limit will be impossible to be overcome in future nano-devices. 

\subsection{The intermediate Landauer's erasure principle }
\label{inter}

The second type of relation between the erasure of information and entropic and energetic changes can be obtained by assuming that the final macroscopic environments are identical at the macroscopic level: 
\begin{itemize}
\item \textbf{Condition C2}: ENVIRONMENTS WITH IDENTICAL FINAL MACROSCOPIC PROPERTIES.  For two
different operations involved in a gate with two initial environment states 
which have macroscopically identical properties at the initial time
(e.g. $\textbf{E}_{\textbf{1}\to \textbf{0}}(t_i)=\textbf{E}_{\textbf{0} \to \textbf{0}}(t_i)$), 
the two final environment states also have identical macroscopic properties (e.g. $\textbf{E}_{\textbf{1}\to \textbf{0}}(t_f)=\textbf{E}_{\textbf{0} \to \textbf{0}}(t_f)$) at the final time \cite{footnote2,footnote2bis}.  
\end{itemize}

Notice that we are not imposing that the initial environment state is macroscopically identical to the final environment state in a given operation (we have shown in the previous subsection that this is impossible for an erasure gate), but only that the two final environment states of the different operations involved in a gate are macroscopically identical. 

\begin{figure}[ht]
\centering
\includegraphics[width=0.6\linewidth]{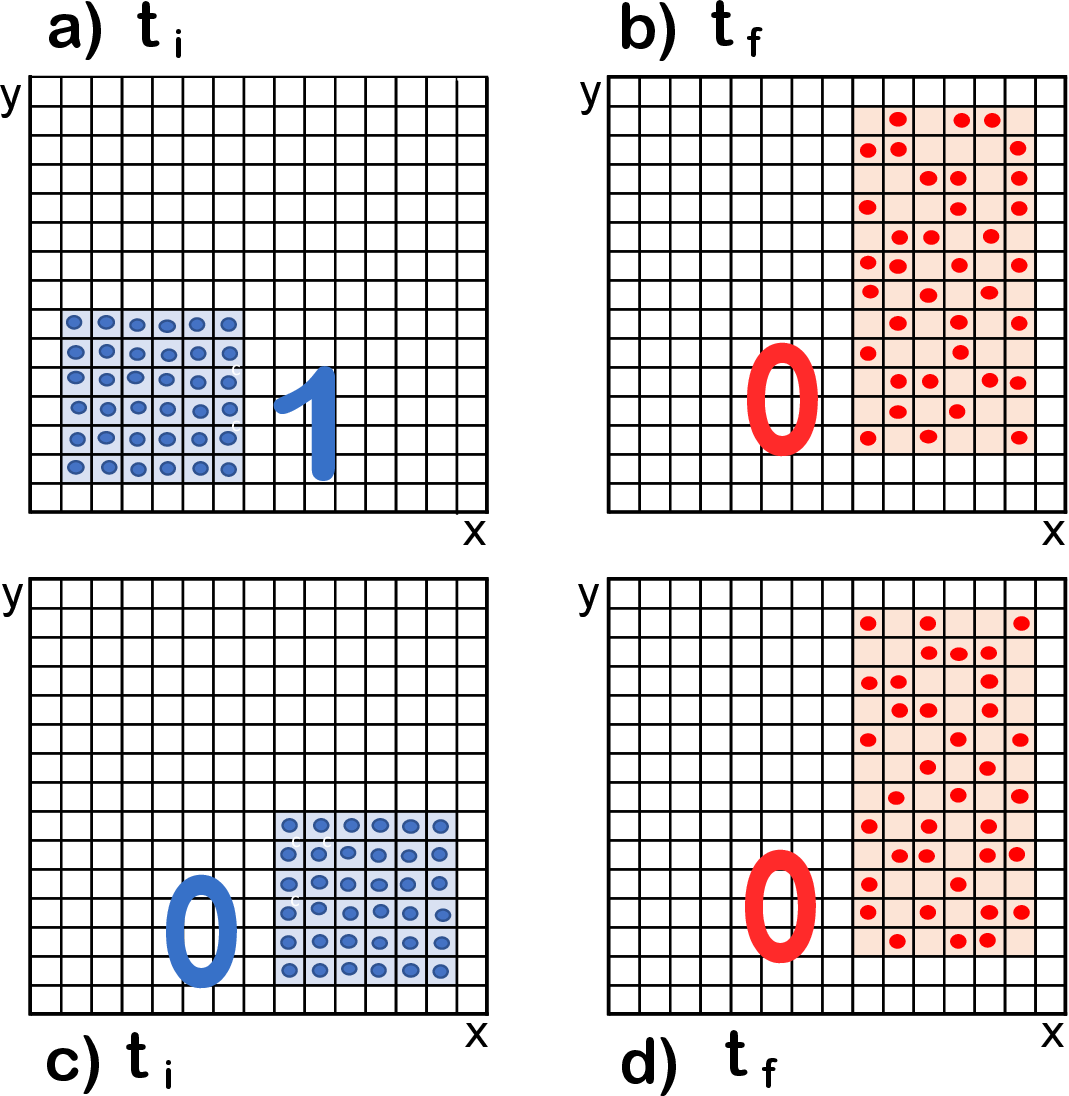}
\caption{Schematic representation of the initial (left panels) and final (right panels) microscopic states (dark blue and red solid circles) and the volumes of macrostates (light blue and red regions) in the system $x$ plus environment $y$ phase space. The upper panels correspond to the operation $\textbf{1}\to\textbf{0}$ and the lower panels to $\textbf{0}\to\textbf{0}$.  The macroscopic property of the system $\textbf{1}$ means being on the left of the $x$ axis, while the macroscopic property of the system $\textbf{0}$ means being on the right of $x$ axis. The initial macroscopic environment properties (in the left panels) are identical $\textbf{E}_{\textbf{1}\to \textbf{0}}(t_i) = \textbf{E}_{\textbf{0} \to \textbf{0}}(t_i)$. The macroscopic properties of the environment at the final time (in the right panels) are also identical $\textbf{E}_{\textbf{1}\to \textbf{0}}(t_f) = \textbf{E}_{\textbf{0} \to \textbf{0}}(t_f)$, in the sense that their microscopic differences are not seen in their macroscopic properties. Even though the gate is logically irreversible, it satisfies the time-reversibility of microscopic laws. The relevant point is that condition \textbf{C2} shows that each operation of the erasure process is done with an increase in entropy: the phase-space volumes (entropies) of the initial macroscopic states (light blue regions in the left panels) are half of the phase-space volumes (entropies) of the final macroscopic states (red blue regions in the right panels). }
\label{schemeinter}
\end{figure}

We analyze again an erasure gate \cite{footnote7} in Fig. \ref{schemeinter} with condition \textbf{C2}. The initial logical states $\textbf{1}$ in Fig. \ref{schemeinter}(a)  and $\textbf{0}$ in Fig. \ref{schemeinter}(c) are different macroscopically (being on the left and on the right respectively), while having the same environment macroscopic states, $E_{\textbf{1}\to\textbf{0}}(t_i)= E_{\textbf{0}\to\textbf{0}}(t_i)$. The final logical states $\textbf{0}$ in Fig. \ref{schemeinter}(b)  and $\textbf{0}$ in Fig. \ref{schemeinter}(d) are macroscopically identical (being on the right). Interestingly, we cannot distinguish macroscopically the final environment macroscopic states, $E_{\textbf{1}\to\textbf{0}}(t_f) = E_{\textbf{0}\to\textbf{0}}(t_f)$ , as seen in the light red regions in Fig. \ref{schemeinter}(b) and (d). If we look microscopically at Fig. \ref{schemeinter}, we see that all the microscopic points (solid red points) in the phase space $\Gamma$ satisfy the time-reversibility imposed by the condition in \ref{empty}, i.e. the solid red point of Fig. \ref{schemeinter}(b) never overlap with the solid red points of Fig. \ref{schemeinter}(d). As we have repetitively stressed, a gate which is physically time-reversible (at the microscopic level) can be logically irreversible (at the macroscopic level). Notice that the distinguishability (or indistinguishability) between two final macroscopic states can have an objective definition, for a example, by imposing a minimum (or maximum) phase space distance between any two final microscopic states belonging to different macroscopic states.  

\begin{proposition}
The erasure of information with a gate satisfying condition \textbf{C2} implies a minimum entropy cost $\Delta S=k_B\ln 2$. 

We define $V_{\textbf{1}\to \textbf{0}}(t_i)$ as the phase-space volume of the  initial macrostate 
$X_{\textbf{1}\to \textbf{0}}(t_i), Y_{\textbf{1}\to \textbf{0}}(t_i)$. Identically for $X_{\textbf{0}\to \textbf{0}}(t_i), Y_{\textbf{0}\to \textbf{0}}(t_i)$, we define $V_{\textbf{0}\to \textbf{0}}(t_i)$. From the proposition in \ref{vb}, 
we now have $v_{\textbf{1}\to \textbf{0}}(t_f)=V_{\textbf{1}\to \textbf{0}}(t_i)$ and 
$v_{\textbf{0}\to \textbf{0}}(t_f)=V_{\textbf{0}\to \textbf{0}}(t_i)$. How can we achieve  condition \textbf{C2} if $v_{\textbf{1}\to \textbf{0}}(t_f) \cap v_{\textbf{0}\to \textbf{0}}(t_f)=\emptyset$ ? The answer is accommodating $v_{\textbf{1}\to \textbf{0}}(t_f)$ and $v_{\textbf{0}\to \textbf{0}}(t_f)$, both, as microstates states belonging to the macrostate $E_{\textbf{1}\to\textbf{0}}(t_f)= E_{\textbf{0}\to\textbf{0}}(t_f)$. To satisfy condition \textbf{C2}, we assume $V_{\textbf{1}\to \textbf{0}}(t_f)=V_{\textbf{0}\to \textbf{0}}(t_f)$ so that $V_{\textbf{1}\to \textbf{0}}(t_f)=V_{\textbf{0}\to \textbf{0}}(t_f)=v_{\textbf{1}\to \textbf{0}}(t_f)+v_{\textbf{0}\to \textbf{0}}(t_f)=V_{\textbf{1}\to \textbf{0}}(t_i)+V_{\textbf{0}\to \textbf{0}}(t_i)$. Assuming that both initial phase-space macroscopic volumes are identical \cite{footnote4}, $V_0=V_{\textbf{1}\to \textbf{0}}(t_i)=V_{\textbf{0}\to \textbf{0}}(t_i)$, we get,
\begin{equation}
\label{Dentropy2}
\Delta S=\frac {1}{2}\Delta S_{\textbf{0}\to \textbf{0}}+\frac {1}{2}\Delta S_{\textbf{1}\to \textbf{0}}=2\frac {1}{2} \left( k_B \ln (2 V_0)-k_B \ln (V_0) \right) =k_B\ln 2.
\end{equation}
We have assumed equal {\it a priori} probabilities for the $\textbf{1}\to\textbf{0}$ and $\textbf{0}\to\textbf{0}$ operations. The reason why the entropy increase is minimal is because  we can imaging $v_{\textbf{1}\to \textbf{0}}(t_f) \cup v_{\textbf{0}\to \textbf{0}}(t_f)$ smaller than the final macrostate, but $v_{\textbf{1}\to \textbf{0}}(t_f) \cup v_{\textbf{0}\to \textbf{0}}(t_f)$ cannot be larger than the final macroscopic state. Again the fundamental microscopic proposition \ref{empty} is satisfied and a Hamiltonian $H_{\textbf{C2}}(x,y)$ is possible \cite{footnote3bis}. 
\end{proposition}

The proof is just a consequence that the number of microstates of the final macrostate is not equal to the number of microstates of the image of the initial macrostate, as seen in Fig. \ref{schemeinter}. This result was already indicated by Landauer himself \cite{landauer1961irreversibility}. Notice, however, that we have made no reference to thermodynamic equilibrium at all in the present development (just counting the number of microscopic states that satisfy a macroscopic property). For this reason, we refer to the result \eref{Dentropy2} as the weak Landauer's erasure principle, because it is more general than the original Landauer limit which implicitly assumed that all the entropy increase was due to a production of heat. In this regard, Bennett wrote \cite{bennett2003notes}  explicitly: \emph{"Typically the entropy increase takes the form of energy imported into the computer, converted to heat, and dissipated into the environment, but it need not be, since entropy can be exported in other ways, for example by randomizing configuration degrees of freedom in the environment."}.

The main conclusion of this subsection is that the increase in entropy can be translated into other types of entropies different from thermodynamic entropy. We note that the same conclusion was reached by the works of Vaccaro and Barnett \cite{barnett2013beyond,vaccaro2011information}. They explicitly generalized the Landauer's erasure principle to new scenarios showing that the costs of erasure depend on the nature of the gate and of the environment with which it is coupled. Their papers were inspired by the enlightening previous work of Jaynes \cite{jaynes1965gibbs} that introduced the concept of the \textit{generalized second law} instead of the usually called \textit{second law of thermodynamics}, to emphasize that the concept of entropy (as a way of counting how many microstates belong to  a given macrostate, as we have done here) does not belong to (equilibrium) thermodynamics only, but can be applied to any system where macroscopic properties matter.  We emphasize that, after accepting that the result $\Delta S=k\ln 2$ has, in general, nothing to do with heat or temperature, new type of gates can be envisioned by looking for new types of entropy different from thermodynamic entropy converted into heat. Such new possibilities will violate the original Landauer's erasure principle in terms of heat and temperature, without violating \eref{Dentropy2} when \textbf{C2} is assumed.

\subsection{The strong Landauer's erasure principle}
\label{strong}

The strong relation between manipulation of information and entropy change leads to the original Landauer's erasure principle. To arrive to it, we invoke the following condition on the final state of environment $E_B$:
\begin{itemize}
\item \textbf{Condition C3}: MACROSCOPICALLY IDENTICAL FINAL THERMAL ENVIRONMENTS.  The final states of environment of different processes of a gate (e.g. $Y_{\textbf{1}\to \textbf{0}}(t_f)$ and $Y_{\textbf{0} \to \textbf{0}}(t_f)$) are described by the same thermal bath \cite{footnote2,footnote2bis}.
\end{itemize}
This condition should be understood as a supplement to \textbf{C2}, i.e. in 
condition \textbf{C3} we assume that condition \textbf{C2} is already satisfied. We are not only imposing that the final states of environment are macroscopically identical, but also that the final states of environment can be described by a state in thermodynamic equilibrium with a well defined temperature $T$.

\begin{proposition}
For an erasure gate satisfying \textbf{C3} (which implies satisfying \textbf{C2} too), the erasure of information implies an increment of heat given by $\Delta Q= k{T}\ln 2$ in the final environments.

For an environment in thermodynamic equilibrium, it is well known that the increment of heat $\Delta Q$ is related to the increment of entropy $\Delta S$ through the thermodynamic relation $\Delta Q = T\Delta S$. Hence, since the increment of entropy is given by \eref{Dentropy2}, we finally have
\begin{equation}
\Delta Q= k{T}\ln 2 .
\label{res2} 
\end{equation}
Here, from the macroscopic property $T$, it is easy to understand how the conditions $Y_{\textbf{1}\to \textbf{0}}(t_f) \cap Y_{\textbf{0}\to \textbf{0}}(t_f) =\emptyset$ (imposed by the proposition \ref{empty}) and $E_{\textbf{0}\to \textbf{0},B}=E_{\textbf{1}\to\textbf{0},B}$ can be satisfied simultaneously. The first refers to microscopic variables (particle positions and momenta) in the phase space, while the second refers to the macroscopic temperature.
In statistical mechanics, there are many different microscopic states corresponding to the same temperature. Again a Hamiltonian $H_{\textbf{C3}}(x,y)$ is possible.
\end{proposition}
  Expression \eqref{res2}  is exactly the original Landauer's erasure principle \cite{landauer1961irreversibility}, which we call the strong Landauer's erasure principle to be distinguished from the previous weak and intermediate ones.  The universality of the strong Landauer's erasure principle in \eref{res2} is based on the assumption that all final states of environments are indeed thermal baths (condition \textbf{C3}). Following the arguments in previous sections and in the next section, the condition \textbf{C3} is a good approximation for many real environments in Nature, but not necessarily valid for all of them (especially if we deal with very fast computations). 

At this point, the reader can wonder why do we insist in the failure of the strong (original) Landauer's erasure 
principle when its limit has been validated by several relevant experiments 
\cite{berut2012experimental,hong2016experimental,yan2018single,orlov2012experimental,berut2015information,berut2013detailed,
ribezzi2019large}, as indicated in Fig. (\ref{References})? All these experiments 
\cite{berut2012experimental,hong2016experimental,yan2018single,orlov2012experimental,berut2015information,berut2013detailed,
ribezzi2019large} have carefully make an effort to ensure that the environment is in thermal equilibrium. 
Then, for thermal environments, the strong Landauer's principle is a universal result. 
Loosely speaking, the experiments are designed to explain the strong Landauer's erasure principle, 
rather than the other way around. In fact, the mentioned experiments have been developed imposing adiabatic conditions 
on the performance of the erasure processes which justify that the environment can be treated as a thermal bath. 
In this regard, the physical transitions seen as left and right distribution of particles in Fig. \ref{schemeinter} cannot be done instantaneously. 
They require some time to thermalize, to change form two distinguishable macrostates to two indistinguishable macrostates. 
Therefore, it seems obvious that in the race for faster computing devices, at some point, the assumption that the environments of an electron devices are always thermalized will not be accurate enough because the reservoir will not have enough  time to thermalize. This very point is in fact what we will discuss in the next section, taking profit of the vast literature on thermalization (or equilibration) in closed quantum systems. 

\section{Can the previous results be extended to the quantum regime?}
\label{quantum}

In this paper, we have shown that the original (strong) Landauer's erasure principle cannot be considered a universal result because it is not true that only thermal reservoirs are available for computations.  The key element in our discussion is the fact that it is possible to envision final environments for the $\textbf{1}\to \textbf{0}$ and $\textbf{0}\to \textbf{0}$ operations with different macroscopic environment properties. But, can we generalize these results to the quantum regime?  Below we provide arguments to justify that it is reasonable to expect that, what we have explicitly demonstrated to be valid for classical erasure gates, is also valid for quantum ones. We note that it is far from the scope of this paper to provide such rigorous quantum extension, here, we only give qualitative evidence of that.   

In the quantum regime, the difference between microscopic and macroscopic levels of description is even 
more important than in classical physics. Microscopic quantum laws seem to be very different from the microscopic classical laws. There is still a strong disagreement in the scientific community on how to define a quantum microscopic state (if it exists at all). In other words, the definition of microscopic states is a rather subtle and controversial issue, because it highly depends on the interpretation of quantum mechanics,
on which there is no consensus among physicists \cite{nik_myth}. A  straightforward demonstration that the developments 
done in Secs. \ref{definition} and \ref{three} can be extended into the quantum regime will be done in the appendix A  
(after selecting a proper interpretation of quantum mechanics). Fortunately, a simple understanding of why condition 
\textbf{C3} is not universal in the quantum regime, and why there is a plenty of room to design erasure gates with 
conditions \textbf{C2} and even \textbf{C1}, can be formulated in an (more or less) interpretation-neutral manner 
(in terms of expectations values) by reusing the recent advances on the process of thermalization of closed quantum 
systems \cite{reviewbo,reviewcoldatoms,reviewclosesystem,equilibration,quantumsimulation,laser,1DBosegases3}. 

Let us suppose that the two operations of the erasure gate are defined by the wave functions $\Psi_{\textbf{1}}(x,y,0)$ 
for the input logical state $\textbf{1}$  and $\Psi_{\textbf{0}}(x,y,0)$ for the input logical state $\textbf{0}$. 
Notice that we are using the variables $x$ and $y$ in the quantum regime as the degrees of freedom of the positions 
of the system and the positions of the environment, respectively. In this sense, $x,y$ represent a point in the 
configuration space, while $x,y$ represented a point in the phase-space in the classical regime. 
The use of the same notation will simplify the comparison of classical and quantum microstates done in the appendix A \cite{footnote10}.

Since the total Hamiltonian $H(x,y)$ is time-independent, the pure states $\Psi_{\textbf{1}}(x,y,0)$ and 
$\Psi_{\textbf{0}}(x,y,0)$ can be described at all times by a unitary evolution 
$\lvert\Psi_{\alpha}(t)\rangle=\sum_n c_{n,{\alpha}} e^{-i E_n t/\hbar} \lvert n\rangle$, with ${\alpha}=\{\textbf{1},\textbf{0}\}$ 
indicating the initial logical state. The ket $\lvert n \rangle$ is an energy eigenstate of the global Hamiltonian $H(x,y)$ mentioned 
in \sref{definition} with eigenvalue $E_n$. Here $c_{n,\alpha}=\langle n\lvert\Psi_{\alpha}(0)\rangle$, 
which depends on the initial wave function, keeps memory of the initial conditions. The density matrix in the energy representation of such global states can be written as
\begin{eqnarray}
\hat \rho_{\alpha}(t)=\lvert\Psi_{\alpha}(t)\rangle \langle \Psi_{\alpha}(t)\rvert=c_{m,\alpha} c_{n,\alpha}^*e^{i(E_n-E_m)t/\hbar} \lvert m\rangle \langle n\rvert=\rho_{\alpha,m,n}(t) \lvert m\rangle \langle n\rvert .
\label{rho1}
\end{eqnarray}
The diagonal elements of the density matrix $\rho_{\alpha,n,n}=\lvert c_{n,\alpha}\rvert^2$ are called \textit{populations}. 
They \textit{forget} the phase of  $c_{n,\alpha}$ and they are time-independent. On the other hand, the off-diagonal elements 
$\rho_{\alpha,m,n}(t)=c_{m,\alpha} c_{n,\alpha}^*e^{i(E_n-E_m)t/\hbar}$ are  called \textit{coherences}. 
They are time-dependent and they quantify the \textit{coherence} between the eigenstates $\lvert n\rangle$ and $\lvert m \rangle$ 
by \textit{keeping memory} of the phase of $c_{m,\alpha} c_{n,\alpha}^*$. See \cite{footnote11} for a discussion of the initial energies. 

By construction of an erasure gate, at the final time $t_f$, the macroscopic properties linked to the system are identical so that we can identify such macroscopic properties of both quantum states with same final logical sate $\textbf{0}$. Then, we assume that a macroscopic property of the environment can be defined from an expectation value \cite{destefani2023assessing,footnote5} of a arbitrary observable $\hat A$ of the environment that can be written as:
\begin{eqnarray}
\langle A \rangle_{\hat \rho_{\alpha}(t)} = \sum_n \lvert c_{n,\alpha}\rvert^2 A_{n,n}+\sum_{n,m\neq n} c_{m,\alpha} c_{n,\alpha}^* A_{m,n} e^{i(E_n-E_m)t/\hbar} ,
\label{rho2}
\end{eqnarray}
where $A_{m,n}=\langle m \lvert \hat A \lvert \lvert n \rangle$. Thus, the discussion on whether 
$\langle A \rangle_{\hat \rho_{\textbf{1}}(t)}  \neq \langle A \rangle_{\hat \rho_{\textbf{0}}(t)}$ or 
$\langle A \rangle_{\hat \rho_{\textbf{1}}(t)} \approx \langle A \rangle_{\hat \rho_{\textbf{0}}(t)}$ 
is a discussion on whether the off-diagonals elements of the density matrix $c_{m,\alpha} c_{n,\alpha}^*e^{i(E_n-E_m)t/\hbar}$ 
(that \textit{keep the memory} of the initial state) are relevant in the evaluation of \eqref{rho2}. 

But such issues have been clarified during the last years in theoretical and experimental works on thermalization of closed quantum systems. In principle, the second term of the right hand side of \eqref{rho2} is a quasi-periodic function different from zero. There are experiments, for example in ultracold quantum gases trapped in ultrahigh vacuum by means of (up to a good approximation) conservative potentials \cite{bloch2008many,cazalilla2011one}, that can be considered to be of the type of systems described above, 
with off-diagonal elements always relevant. The near unitary dynamics of such systems has been observed in beautiful experiments on collapse and revival phenomena of bosonic \cite{greiner2002collapse,will2013coherent} and fermionic \cite{will2015observation} 
fields, without the relaxation phenomena predicted with traditional ensembles of statistical mechanics 
\cite{1DBosegases1,langen2015experimental}. Thus, as we have argued along the paper, there are computing scenarios where 
the condition \textbf{C1} that environments are macroscopically distinguishable is physically viable. 
In fact, all these works on quantum thermalization of closed systems have been motivated to understand the recent constructions 
of several prototypes of the so-called quantum simulations where the behavior of a quantum system, 
which cannot be solved numerically due to the many-body problem of the Schrodinger equation, 
is empirically realized in the laboratory by studying the evolution of another controlled quantum system that mimics 
the first one. Obviously, in such (analog) quantum simulations, and also in (digital) quantum computations dealing with qubits,  
the need of controlled (non-thermal) environments is mandatory for minimization of decoherence phenomena. 

It is true that a system satisfying condition \textbf{C1} requires an important technological effort on engineering the behavior of the environment. In fact, there are other quantum closed systems that do thermalize and such processes have been reasonably well-understood too. At the initial time $t=0$, it is clear that $\langle A \rangle_{\hat \rho_{\textbf{1}}(0)}  \neq \langle A \rangle_{\hat \rho_{\textbf{0}}(0)}$ because we start from macroscopically different states. But, after some time we can find $\langle A \rangle_{\hat \rho_{\textbf{1}}(0)}  \approx \langle A \rangle_{\hat \rho_{\textbf{0}}(0)}$ if the off-diagonal terms become irrelevant. A simple argument can clarify the need for a delay to reach equilibration in a closed quantum system.  Even if none of the terms $c_{m,\alpha}$, $c_{n,\alpha}^*$ and $ A_{m,n}$ are exactly zero at any time, it is possible to envision a scenario in which the whole sum of the right hand side of \eqref{rho2} is close to zero because the off-diagonals terms cancel each other due to adding of effectively \textit{random} complex numbers. However, such \textit{randomization} requires some time, which is called \textit{equilibration} time $t_{eq}$ in the literature. Then, the time evolution of $\langle A \rangle$ after the equilibration time  $t>t_{eq}$, when the off-diagonal elements of the density matrix are no longer relevant, can be described by a time-independent diagonal density matrix in the energy representation $\hat \rho_{\rm diag}=\sum_n \lvert \langle n\lvert c_n\rangle \rvert^2 \lvert n\rangle \lvert \langle n \rvert$. In the literature, it is said that a quantum system suffers equilibration when the expectation value in \eqref{rho2} satisfies 
$\text{tr}\{ \hat A \hat {\rho} \}\approx \text{tr}\{ \hat A \hat \rho_{\rm diag}\}$  
for the overwhelming majority of times (allowing for some sporadic revivals) larger than the equilibration time $t_{eq}$. 
Notice that the diagonal matrix, after that, does not yet need to be a (micro-canonical, canonical or grand-canonical) thermal 
density matrix. In any case, we get macroscopically identical properties of the environments, 
$\langle A \rangle_{\hat \rho_{\textbf{1}}(t)} \approx \langle A \rangle_{\hat \rho_{\textbf{0}}(t)}$. 
This corresponds to condition \textbf{C2}, where the environments are indistinguishable but not thermal yet. 
Again, a lot of experimental work on such quantum equilibration scenarios is present in the literature 
\cite{reviewcoldatoms,reviewclosesystem,equilibration,quantumsimulation}. When  $\hat \rho_{\rm diag}$ is roughly equal 
to the micro-canonical (canonical or grand-canonical in open systems) density matrix, then the quantum system is said 
to be thermalized. This corresponds to condition \textbf{C3}. 

In the literature \cite{reviewcoldatoms,reviewclosesystem,equilibration,quantumsimulation,laser,1DBosegases3}, 
one can find equilibration times $t_{eq}$ ranging from  few femtoseconds to picoseconds, depending on the details and 
complexity of the systems at hand. If we define $\theta_{n,\alpha}$ as the phase of $c_{n,\alpha}$, a simple 
(but not rigorous) estimation of $t_{eq}$ can be obtained by noting that at $t=0$ 
all phases of the off-diagonal elements (coherences) satisfy 
$e^{i(\theta_{n,\alpha}-\theta_{m,\alpha})}e^{i(E_n-E_m)0/\hbar}=e^{i(\theta_{n,\alpha}-\theta_{m,\alpha})}$, 
so that all phases of the off-diagonal elements together perfectly keep memory of the initial sate $\Psi_{\alpha}(x,y,0)$. 
To forget such memory, we require that the sum of all coherences in \eqref{rho2} vanishes after an equilibration 
time $t_{eq}$. If such equilibration occurs, all relevant phases $e^{i(E_n-E_m)t_{eq}/\hbar}$ have to reach a value equal 
or larger than $2\pi$ to ensure that $e^{i(\theta_{n,\alpha}-\theta_{m,\alpha})}e^{i(E_n-E_m)t_{eq}/\hbar}$ are randomly distributed.
If we define $\Delta E_{eq}={\rm min}(E_n-E_m)$ of all relevant energies of the system, a simple estimate of the equilibration time is given by
\begin{eqnarray}
t_{eq} \approx \frac{h}{\Delta E_{eq}} ,
\label{teq}
\end{eqnarray}
where $h$ is the Planck constant. For a reservoir of length $L=100$ nm, with a parabolic relation between energy and momentum, we can estimate a minimal energy gap between energy eigenstates equal to  $\Delta E\approx 10^{-3}$ or $10^{-4}$ eV.  If we use $\Delta E\approx 10^{-4}$ eV in expression \eqref{teq}, we get an approximate value of the equilibration time $t_{eq} \approx 1$ ps. Even though the formula \eqref{teq} is not rigorous at all, it clarifies that process of thermalization, in our case changing from different macroscopic properties $\langle A \rangle_{\hat \rho_{\textbf{1}}(0)}  \neq \langle A \rangle_{\hat \rho_{\textbf{0}}(0)}$ to identical macroscopic properties $\langle A \rangle_{\hat \rho_{\textbf{1}}(t_{eq})}  = \langle A \rangle_{\hat \rho_{\textbf{0}}(t_{eq})}$, cannot be instantaneous but  requires a time to occur. This conclusion can alternatively be reached from the definitions of Markovian and non-Markovian open quantum systems \cite{opensystem}. An open quantum system interacting with an environment is, in principle, a non-Markovian system. The evolution of the system (together with the environment) can only be considered Markovian if we consider the evolution in (coarse-grained) time steps larger than the time interval needed for the environment to relax. Thus, the transition from non-Markovian to Markovian relaxation time also requires a time related to the relaxation of the environment. In conclusion, even in typical environments where the assumption of thermalization is reasonable, we cannot have an instantaneous thermalization process. This delay in the thermalization provides an unquestionable limit on the speed of computations to satisfy the basic assumption of the strong (original) Landauer limit. This limit is also shown in Fig. \ref{historicaldissipation}. Beyond THz frequencies, the assumption that environments are always macroscopically identical to a thermal bath is not admissible and the original Landauer dissipation seems not applicable.

\section{Conclusions}
\label{conclusions}

After more than 60 years, the Landauer's erasure principle is still accompanied by controversies. In this regard, Landauer himself wrote \cite{landauer1994zig}. 
{\it "The path to understanding in science is often difficult. If it were otherwise, we would not be needed. 
This field [fundamental physical limits of information handling], 
however, seems to have suffered from an unusually convoluted path."} 
What we find especially unfortunate during the recent developments in this field is linking the result of the dissipation 
in computing gates to equilibrium thermodynamics \cite{footnote9}. This link is unfortunate because it is not only 
unnecessary (as we have seen in our paper), but it has the undesired effect of unnecessarily limiting the imagination 
of many researchers. An exception that has overcome this limitation has recently been published in Ref.  \cite{work}, where erasure gates using squeezed thermal environments are proposed.  

Thus, at first sight, it seems that any attempt to discuss possible extensions of the Landauer's erasure principle beyond thermodynamic equilibrium requires the flexible tools of non-equilibrium thermodynamics. Such non-equilibrium tools will certainly still require some notion of \textit{equilibrium} to be able to define what is heat, work, etc. As we mentioned, this is the typical path followed for most investigations on Landauer's extensions. But, this is not the path we have followed in our paper. Can we use a description of the erasure process based exclusively on the mechanical (not thermodynamic) laws of physics? Yes, of course. An erasure gate is, at the end of the day, a physical system whose performance follows the fundamental microscopic laws of physics. As an example, in appendix B, we have shown a toy model of an erasure gate whose performance during several repetitions is evaluated by numerically solving the fundamental microscopic laws that simultaneously govern the degrees of freedom of the system and the environment. 

The reader can (erroneously) argue that we have used some thermodynamics concepts, not only microscopic laws, along the paper because we have included entropy argumentations. We have only used a definition of entropy as the number of microstates that are present in a given macrostate. By construction, such concept of entropy is perfectly adequate in a microscopic description of any system (independently on whether it is used in thermodynamic discussions too). It only requires the proper definition of a macroscopic state in terms of microscopic states, as we have done in Section \ref{definition}. Then, of course, in subsection \ref{strong} we have invoked the equilibrium thermodynamics concepts of heat and temperature, but only to reach the original Landauer formulation, which is nothing but a special case of our general formulation.  

The main advantage (and drawback) of our paper is that it uses classical microscopic physics. As such, it provides a mathematically simple and physically rigorous understanding of the three types of Landauer's erasure principle. But, strictly speaking, the results of this paper have not been demonstrated to be valid in quantum scenarios. A rigorous quantum extension of the classical microscopic explanation presented here is far from the scope of this paper. The main reason is because there is still a strong disagreement in the scientific community on how to define a quantum microscopic state (if it exists at all). In fact, even the wave function (linked to any definition of a microstate) is under a lively debate now (does it represent only epistemic knowledge about the outcomes of future measurements? or, is it something ontologically real ?) \cite{nik_myth}. Even, it is not clear if the wave function is enough to define a microscopic state, since it is also argued that present quantum theory has to be understood as something emergent; as an average description of an underlying more complicated quantum dynamics (with additional microscopic variables) \cite{nik_myth}. Despite this poor understanding of what quantum microscopic states are, in Section \ref{quantum} we have provided some quite general evidences that it is reasonable to expect that the classical results presented here do also apply in a quantum regime. Basically, even under the assumption that a quantum environment will effectively reach some type of \textit{equilibrium} (whatever it means), some time will be needed to reach it. In addition, in the appendix A, after selecting a particular interpretation of quantum mechanics, we also provided a natural extension of the classical results of the main text to the quantum regime. 

Finally, let us mention that the strong Landauer's erasure principle has not been relevant yet for practical devices because nowadays 
other larger sources of dissipation are present. It seems reasonable to expect that in the future, when the other sources 
of dissipation disappear,  the strong Landauer's erasure principle will still not be relevant because future computing 
devices will work at frequencies for which the assumption of environment in (classical or quantum)  thermodynamic equilibrium will no longer be valid as shown in the shaded region in Fig. \ref{historicaldissipation}. 

We hope that the present work will help to develop new research avenues 
for engineering computing devices with environments that satisfy condition \textbf{C2} involving entropy change without 
heat dissipation, or even approaching condition \textbf{C1} where the entropy change can be reduced significantly. 

\backmatter

\begin{appendices}

\section{Counting the number of microstates} 

In this appendix A, in the first subsection, we will show that the concept of entropy is a way to quantify the number of microstates that belong to a given macrostate. Then, in the second subsection, we will show that the development done in the manuscript in terms of well-defined trajectories can be extended into the quantum regime in a very simple and natural way by using quantum (Bohmian) trajectories. 

\subsection{ Classical procedure}

We consider a classical system in an experiment described by the trajectory $x^{(j)}(t),y^{(j)}(t)$ that belongs to the macroscopic state $A$. Then, the entropy \cite{shelly} is defined as
\begin{equation}
\label{entropy}
S(t)=k_B \ln\big( V_A\big),
\end{equation}
where $k_B$ is the Boltzmann constant and $V_A$ is the volume in the phase space, definition \ref{va}, for all $M$ points of the phase space that \textit{look macroscopically  similar} to $x^{(j)}(t),y^{(j)}(t)$, that is all the phase-space points of the macroscopic state $A$ of the macroscopic property $\textbf{A}$, according to definition \ref{macro}.

\begin{figure}[ht]
\centering
\includegraphics[width=0.5\linewidth]{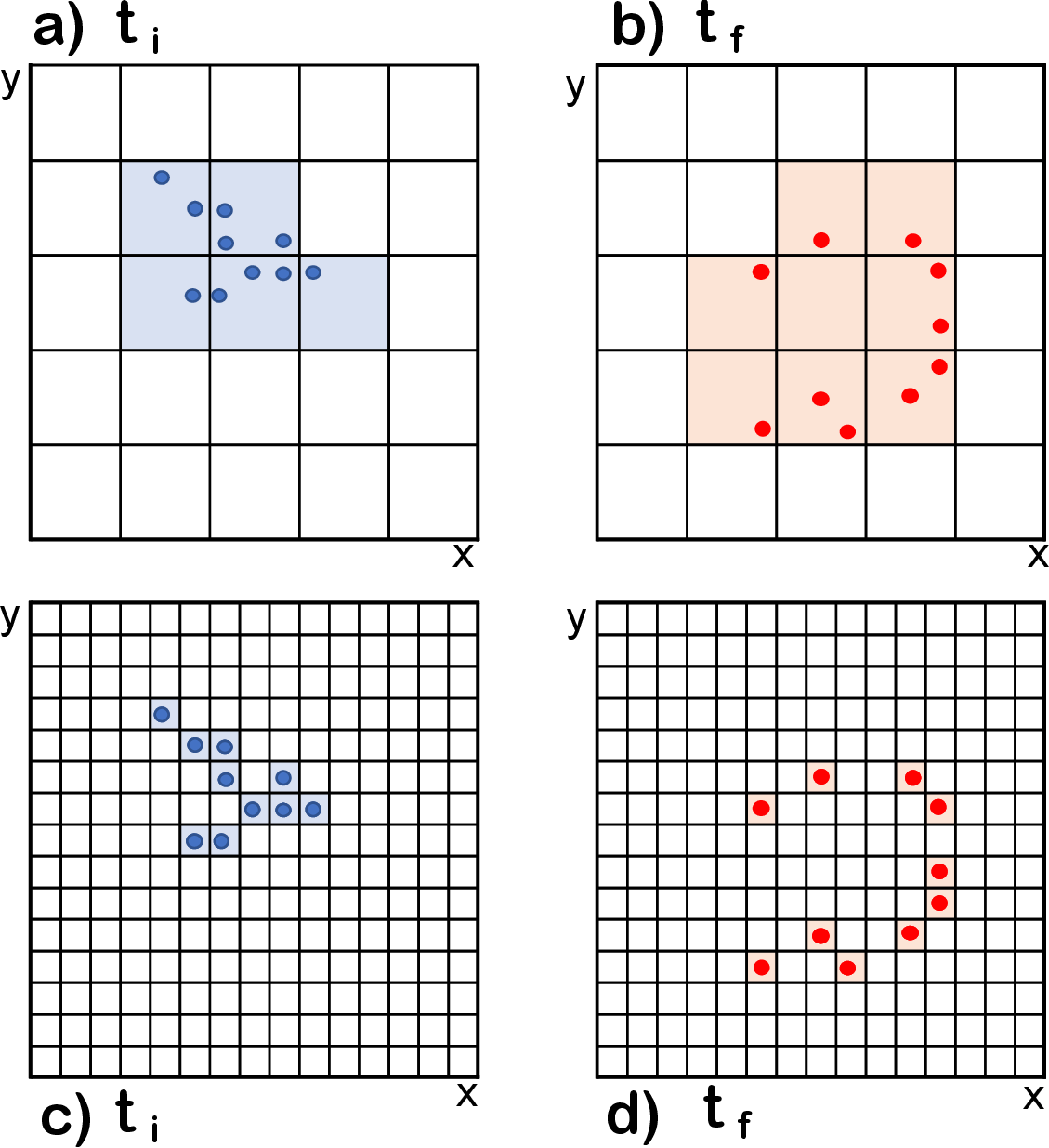}
\caption{Schematic representation of the initial (left panels) and final (right panels) microscopic states 
(dark blue and red solid circles) and the volumes of macrostates (light blue and red regions) 
in the system $x$ plus environment $y$ phase space 
(or in the system $x$ plus environment $y$ configuration space  for a quantum system). The upper and lower panels describe the coarse and fine graining, respectively, of the same initial and final microscopic states, explaining why the volume of a macrostate can be quantified by just counting the number of microscopic states. In the upper panels a large grid is assumed so that the space is divided into cells. In (a) the number of \textit{occupied} cells is $5$ and the number of microstates is $M_A=10$. In (b) the number of  \textit{occupied} cells is $8$ and the number of microstates is $M_A=10$. Since the Liouville theorem ensures that $V_A$ in (a) has to be equal to $v_B$ in (b), from $5 \ne 8$ we see that such a grid does not allow us to identify the volume of the macrostate with the number of microstates. In the lower panels a smaller grid is chosen, by dividing the phase space into smaller cells so that only one microstate (or none) occupies each cell. In (c) and (d) the number of cells is $10$ and the number of microstates is $M_A=10$. For such a grid, identifying the volume of the macrostate with the number of microstates is correct. If more microstates belonging to the macroscopic property $\textbf{A}$ are considered in the discussion, we can use a grid with even smaller cells until we satisfy again the requirement that only one microstate (or none) occupies each cell. }
\label{volume}
\end{figure}

 In Fig. \ref{volume}, we have represented a $L_x \times L_y$ region of the phase space $\Gamma$ and the microscopic points of the macroscopic state $M_A$ at the initial $t_i$ and final $t_f$ times. We have drawn 25 cells of area $\Delta \Gamma=L_x/5\times L_y/5=L_x\times L_y/25$ in Figs. \ref{volume}(a) and (b). It seems from these plots that phase-space volume of $M$ is not proportional to the number of phase-space points (because it also depends on the size of the cells). However, in Figs. \ref{volume}(c) and (d) we select 225 smaller cells with an area $\Delta \Gamma'=L_x/15\times L_y/15=L_x\times L_y/225$ so that each cell accommodates only one (or zero) point. Now, the phase-space volume becomes proportional to the $\Delta \Gamma'$.  Thus, we can define the entropy for the $M_A=10$ points at the initial time $t_i$ as
\begin{equation}
\label{entropy2}
S(t_i)=k_B \ln\big( M_A \Delta \Gamma' \big)=k_B \ln\big( M_A\big)+k_B \ln\big(\Delta \Gamma'\big) .  
\end{equation}
Notice that the use of a smaller area than $\Delta \Gamma'$ in \eqref{entropy2} (for example, $\Delta \Gamma''$ so that each cell still contain zero or one microstate) will only modify the last constant, $k_B \ln\big(\Delta \Gamma'\big) \to k_B \ln\big(\Delta \Gamma''\big)$, which is irrelevant when evaluating the entropy change between two times (as far as both use the same cell grid). Similarly, increasing the number of points will only modify the same irrelevant constant (as far as each cell accommodates only one or zero points). Thus, as it is well-known, the entropy linked to the macrostate $A$ can be computed from the number of microstates $M_A$ that belong to such macrostate. 
    
\subsection{Quantum procedure}

As indicated in the manuscript, identifying the microscopic properties of a quantum systems is a 
rather subtle and controversial issue, because it highly depends on interpretation of quantum mechanics,
on which there is no consensus among physicists \cite{nik_myth}. For example, in the standard interpretation of 
quantum mechanics, it is the many-body wave function in the configuration space that provides such microscopic description, 
while other interpretations may postulate additional variables or replace the wave function with an entirely different object. 
Since each theory provides also its equation of motion for its microscopic state, the diversity of quantum theories 
implies no consensus on the \textit{behavior} of the quantum microscopic world. Some interpretations of quantum mechanics 
claim that the quantum laws for such microscopic states are neither deterministic nor time-reversible. 
By contrast, other interpretations claim that the microscopic quantum laws are deterministic and reversible, while 
indeterminism and irreversibility emerge only at the macroscopic level. At this point, a quantum procedure to count 
the number of microstates requires specifying which quantum theory is used to interpret quantum phenomena. 
The authors of this paper believe that the conceptually clearest way to think about quantum physics, in general, 
and on quantum microscopic states, in particular, is to use the Bohmian interpretation of quantum mechanics \cite{bohm1952suggested,oriols_book,nik_IBM,teufel2009bohmian}, in which
the differences between classical and quantum microscopic physics look less radical than in other interpretations. 
We choose in this appendix A this interpretation to explicitly show that all developments done in Secs. 
\ref{definition} and \ref{three} for classical systems can be straightforwardly extended to the quantum regime.

The fundamental elements of the Bohmian theory are the many-body wave function 
$\Psi(x,y,t)$ of a closed quantum system, together with the actual particle positions $x^{(j)}(t)$ of the system and actual 
particle positions $y^{(j)}(t)$ of the environment \cite{bohm1952suggested,nik_IBM,teufel2009bohmian}. 
The evolution in time of such positions $x^{(j)}(t),y^{(j)}(t)$ represents a trajectory in the configuration space. 
The many-body wave function $\Psi(x,y,t)$ guides the particles by determining their 
velocities \cite{bohm1952suggested,oriols_book,nik_IBM}. 
In the laboratory, at the initial time $t=0$, one can \textit{prepare} the wave function $\Psi(x,y,0)$, 
but one cannot \textit{prepare} the initial positions \cite{durr1992quantum}. 
Such initial positions $x^{(j)}(0)$ and $y^{(j)}(0)$ obey the probability distribution $\lvert \Psi(x,y,0)\rvert^2$ 
when the identical experiment is repeated $M\to\infty$ times, where the superindex $j$ labels each experiment 
$j=1,..,M$. Thus, once $x^{(j)}(0)$, $y^{(j)}(0)$ and $\lvert \Psi(x,y,0)\rvert^2$ are fixed, the Bohmian interpretation of quantum 
phenomena is deterministic and time-reversible at the microscopic (ontological) level, 
just like classical physics. However, since we have no direct control over $x^{(j)}(0)$ and $y^{(j)}(0)$, 
the Bohmian results are random at the empirical level \cite{oriols_book,nik_IBM,teufel2009bohmian}. 
As indicated in \sref{quantum}, we are using the variables $x$ and $y$ in the quantum regime as a point 
in the configuration space, rather than as a point in the phase-space as in Secs. \ref{definition} and \ref{three} 
for classical particles. This is the only difference between the quantum (Bohmian) and classical description of microstates. 
We have seen  in the first part of this appendix A that such difference becomes irrelevant when discussing the three types of Landauer's erasure principle 
explained in the manuscript. Therefore, the use of the same notation for quantum and classical versions will help in 
straightforwardly reusing the classical results for quantum systems.

First of all, if we want to count quantum microstates, we have to specify in detail what is the definition 
of a \textit{microscopic state} in the Bohmian theory.  At first sight, it seems that a microscopic state is determined by trajectory 
$x^{(j)}(t),y^{(j)}(t)$ of the $j$-th experiment plus the wave function $\Psi(x,y,t)$ that guides such trajectories. 
But, in fact, the Bohmian theory is a holistic theory in the sense that its applicable to the whole Universe, 
not only to a part of it \cite{nik_IBM,durr1992quantum,teufel2009bohmian}. Therefore, strictly speaking, there is 
just one wave function for describing any process in the Universe. For our practical example of an erasure gate, 
it means that the wave functions $\Psi_{\textbf{1}}(x,y,t)$ for the input logical state $\textbf{1}$  and 
$\Psi_{\textbf{0}}(x,y,t)$ for the input logical state $\textbf{0}$ used in \sref{quantum} can be substituted by a 
unique \textit{big} wave function of a closed system. This \textit{big} wave function appears in a natural way in 
Bohmian mechanics by taking into account the rest of degrees of freedom of the laboratory, labeled here as $z$, 
that determine whether we are a dealing with a initial logical $\textbf{1}$ or $\textbf{0}$. Such \textit{big} wave function 
is written as $\Psi(x,y,z,t)$, and we can define the wave function for  $\textbf{1}$ as the conditional (Bohmian) 
wave functions \cite{oriols_book,durr1992quantum,teufel2009bohmian} of the \textit{big} wave function as 
$\Psi_{\textbf{1}}(x,y,t)\equiv\Psi(x,y,z^{j_1}(t),t)$. Identically, we define  the wave function describing the 
initial logical $\textbf{0}$ as $\Psi_{\textbf{0}}(x,y,t)\equiv \Psi(x,y,z^{j_0}(t),t)$. Here, $z^{j_1}(t)$ 
can be any of the configurations of the positions $z$ of the set-up of the laboratory (excluding the environment and the system) 
that are linked to an initial logical state $\textbf{1}$. Likewise,  $z^{j_0}(t)$ for $\textbf{0}$. The final result is that 
a quantum (Bohmian) microstate is defined just by $x^{(j)}(t),y^{(j)}(t),z^{(j)}(t)$ alone. We do not need to include the wave function 
in our attempt to count quantum microstate because, despite the wave function is a fundamental ontological element of the 
Bohmian theory, it is the same for all experiments of the erasure gate. By including $z^{(j)}(t)$ as part of the degrees of 
freedom of the environment, the quantum (Bohmian) and classical definition of a microstate are almost identical 
(the first is a point in the configuration space, while the second a point in the phase space). Once this small difference 
is accounted for, all definitions done in Secs. \ref{definition} and \ref{three} for counting the number of classical 
microstates can be reused for getting identical conclusions in the quantum regime. For example, the concept of entropy in the quantum 
case just requires reinterpreting the phase-space axes in the figures (like Fig. \ref{volume}) as the axes in the 
configuration space. Such a change of axes is inessential because we have demonstrated in the first part of this appendix A that what matters 
in the discussion of entropy is just the number of microstates that belong to each macrostate. 
Importantly, the non-crossing property of the classical trajectories in the phase-space in proposition \ref{empty} 
is also satisfied by the quantum (Bohmian) trajectories in the configuration space. Since all experiments deal 
with a unique wave function $\Psi(x,y,z,t)$ which is a single-valued function, so only one velocity 
at position $x,y,z$ and time $t$ is possible. This implies that 
Bohmian trajectories cannot cross in the configuration space \cite{oriols_book,teufel2009bohmian}).

\section{Toy model for the weak Landauer's erasure principle}
\label{appb}

In this appendix B, we describe a toy model of an erasure gate that satisfies the weak Landauer's erasure principle (condition \textbf{C1}). The model does not pretend to be realistic, but just exemplify the physical soundness of this weak version. We consider a 2D system where each particle position has a horizontal $r_x$ and vertical $r_y$ location. Notice that, in this appendix, we use $x$ and $y$ as directions in physical space which is different from the meaning assigned to them in main part of the paper. 

\begin{figure}[ht]
\centering
\includegraphics[width=0.6\linewidth]{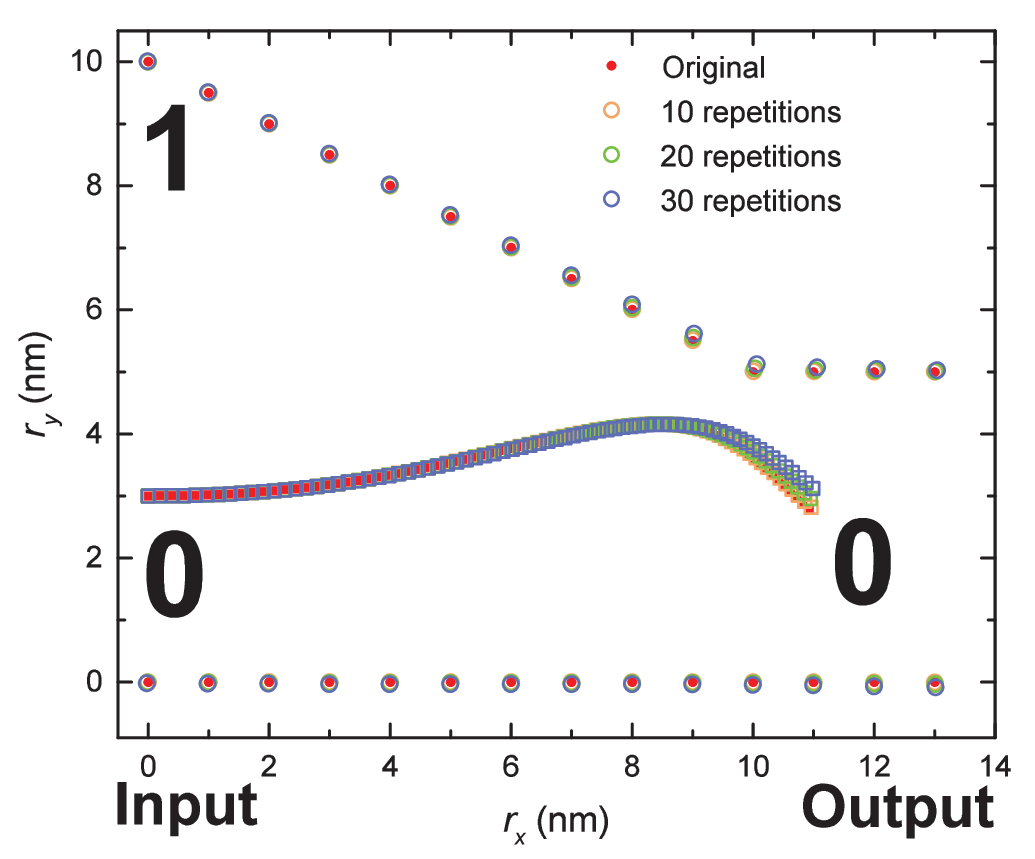}
\caption{Erasure gate (toy model) representing several repetitions of the logical operations $\textbf{0}\to \textbf{0}$. Red solid circles denote original initial positions of the $N_E=28$ particles of the environment during the first operation. Red solid squares denote the original trajectory of the single particle of the system, from the input $r_x=0$ to the output $r_x=11$, plotted every 0.2 fs . During the first $\textbf{0}\to \textbf{0}$ operation, each particle of the environment interacts with the system particle and with the other particles and modifies its position and velocity. When the first system particle leaves the gate, a new system particle enters into the gate, and a repetition of the $\textbf{0}\to \textbf{0}$ logical operation is done with the initial conditions of the environment particles in this second operation equal to their final conditions in the first operation. No reset of the environment is considered after each operation. The initial positions of the particles of the environment at the 10, 20 and 30 repetitions are indicated with empty circles with different colors. The same colors are used to represent the corresponding trajectory of the system at these repetition with empty squares. }
\label{figureab1}
\end{figure}

\begin{figure}[ht]
\centering
\includegraphics[width=0.6\linewidth]{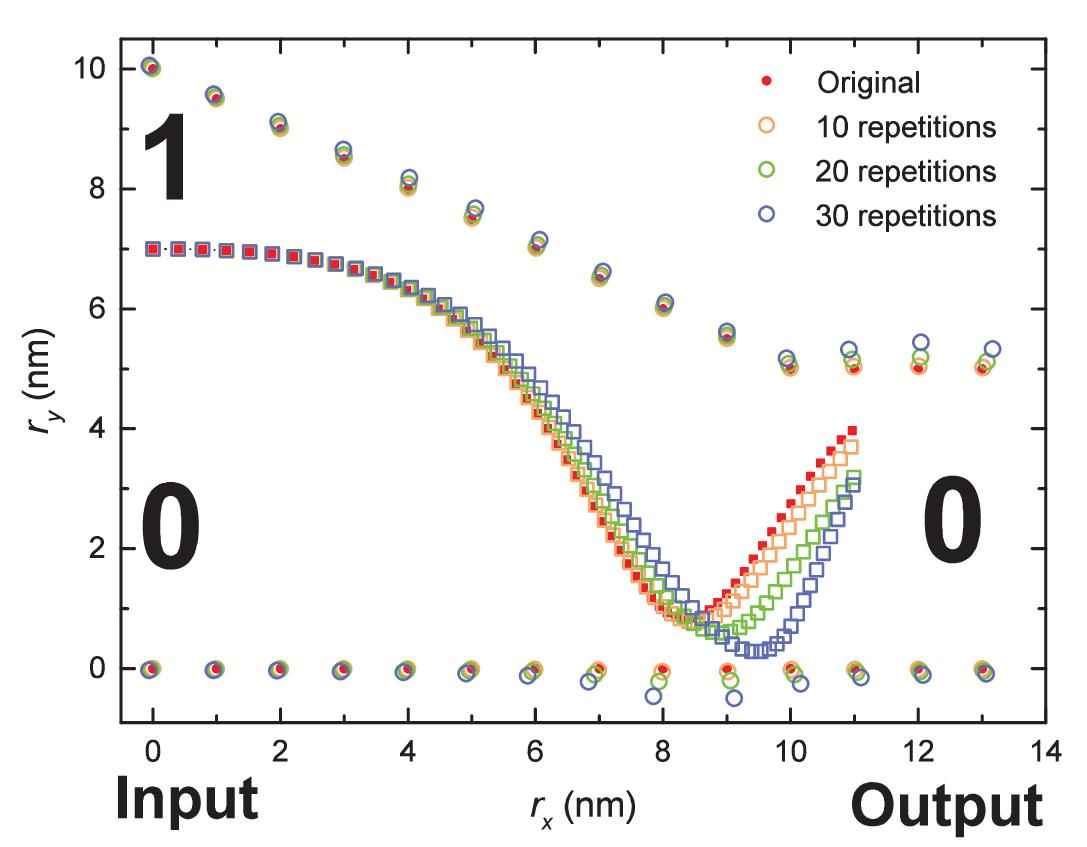}
\caption{Erasure gate (toy model) representing several repetitions of the logical operation $\textbf{1}\to \textbf{0}$. All the details are identical to those of Fig. \ref{figureab1} except for the initial position of the system at each repetition, which is here $\{r_{x,29}(0)=0\; nm,r_{y,29}(0)=7 \; nm\}$, while it was $\{r_{x,29}(0)=0\; nm,r_{y,29}(0)=3 \; nm\}$ in Fig. \ref{figureab1}. }
\label{figureab2}
\end{figure}

The number of particles of the environment, $N_E=28$, can be separated into two sets. The first set is formed by the 14 particles indicated by the red solid circles in the top of Figs. \ref{figureab1} and \ref{figureab2}. The second set of 14 particles is indicated by red solid circles in the bottom of Figs. \ref{figureab1} and \ref{figureab2}.  The first set can be identified as a ``top barrier'' , while the second as a ``bottom barrier''. Both ``barriers'' will guide the system particle from the input toward the output of the gate. In order to minimize the movements of the particles of the environment for the reasons that will become clear later, each particle in the environment has a mass of $m_i=4000 m$ with $m$ the free electron mass and a charge $q_i=0.04\; q$ with $q$ the electron charge (with sign). The initial velocity of the particles of the environment is zero in both directions, $v_{x,i}(0)=0$ m/s and $v_{y,i}(0)=0$ m/s.

In fact, since we are interested in using the same erasure gate several times, there is just one environment ($N_E=28$ particles) involved all the time, but several systems. We consider the system as a single particle that enters into the gate at $r_x=0$ nm encoding an input logical value , travels inside, and exits the gate at $r_x=11$ nm encoding the output logical value. Using another time, the same gate implies using the same environment but a new degree of freedom for the system encoding the new logical information. To avoid  a complicated notation, since only one system particle is interacting with the environment at each operation, we will assume that each system particle is described by a degree of freedom label as $n=29$. The mass of the particle of the system is $m_{29}=0.2m$ and its charge $q_{29}=10 q$.

Then, the dynamics of all particles (environment plus system), at each operation, is determined by the following Hamiltonian.
\begin{eqnarray}
H_{C1}=\sum_{i=1}^{29} \left(\frac {p_{x,i}^2}{2 m_i}+\frac {p_{y,i}^2}{2 m_i}\right)+\frac{1}{2}\sum_{i=1}^{29}\sum_{j=1}^{29}\frac{1}{4\pi \epsilon} \frac{q_i q_j}{\sqrt{(r_{x,i}-r_{x,j})^2+(r_{y,i}-r_{y,j})^2}}
\label{hamapb}
\end{eqnarray}
with $\epsilon$ the vacuum permittivity and $p_{x,i}=v_{x,i}\; m_i$ and $p_{y,i}=v_{y,i}\; m_i$ the momentum components of the $i$-th particle in the horizontal and vertical directions, respectively. Identically, $r_{x,i}$ and $r_{y,i}$ are the position components of the $i$-th particle in the horizontal and vertical directions, respectively. Notice that the ``top barrier'' and ``bottom barrier'' are not external potentials, but just particles interacting with the system particle. The numerical solution of the interacting $N=29$ particles  is done by time-integrating the acceleration, computed from Newton's laws, with a temporal step of $1\cdot10^{-18}$ s. 

In Fig. \ref{figureab1}, apart from the initial positions of the environment in solid red circles, we plot the trajectory of the system particle in solid red squares from the input of the gate ($r_x=0$ nm) till the output ($r_x=11$ nm). The initial state of the system corresponds to the logical $\textbf{0}$ described as  $\{r_{x,29}(0)=0\;nm,r_{y,29}(0)=3 \; nm \}$ and initial velocity $v_{x,29}(0)=2\cdot10^6$ m/s in the horizontal direction and zero $v_{y,29}(0)=0$ m/s in the vertical direction.The particle of the system is initially repelled by the ``bottom barrier'', and later by the ``top barrier''. Finally, at the horizontal position of $r_x=11\;nm$, the system indicates the final logical value $\textbf{0}$. As a consequence of the interactions given by \eqref{hamapb}, the particles of the environment have slightly modified their initial positions and velocities.

As it happens in a real gate, after the first operation, the gate is ready for a second operation. Such second operation happens when another particle of the system is prepared identically to the first system particle, indicating the initial logical $\textbf{0}$. However, now the environment particles are not the solid red circles in Fig. \ref{figureab1}, but the new initial conditions of the particles of the environment in this second operation correspond to the final conditions of the environment after the first operation. The environment of the second operation is different from the environment of the first one. This is exactly what happens in the strong Landauer's erasure principle. Each time an operation takes place, the final environment becomes hotter than before the operation. As far as the hotter final environment is not much different from the initial one, no reset of the environment degrees of freedom is needed in real erasure gates.     

Since we are also interested in avoiding the reset of the environment in our toy model of the weak Landauer's erasure principle, we want an environment that suffers small perturbation. Now, it becomes evident why we select such heavy particles ($m_i=4000 m$) with such small charge ($q_i=0.04\; q$) in the particles of the environment. In Fig. \ref{figureab1} we have plotted in circles of different colors the final positions of the particles of the environment at different repetitions of the operation (without a reset in the environment). We also plot the trajectories of the system with squares and with the same colors that we used for the environments.    

In Fig. \ref{figureab2}, we have repeated the results of In Fig. \ref{figureab1}, but now considering that the system corresponds to a logical $\textbf{1}$ defined as $\{r_{x,29}(0)=0\; nm,r_{y,29}(0)=7 \; nm\}$ with the same initial velocity of the system particle as before: $v_{x,29}(0)=2\cdot10^6$ m/s in the horizontal direction and zero $v_{y,29}(0)=0$ m/s in the vertical direction. The system particle is now initially repelled by the ``top barrier'', and later by the ``bottom barrier''. Finally, at the horizontal position of $r_x=11\; nm$, the system's microscopic state corresponds to the final logical state $\textbf{0}$. As a consequence of the interactions given by \eqref{hamapb}, the particles of the environment in Fig. \ref{figureab2} have a stronger modification of their initial positions and velocities than in Fig. \ref{figureab1}. Roughly speaking, it is more ``difficult'' to convert the initial $\textbf{1}$ into a final $\textbf{0}$ in Fig. \ref{figureab2}, than to keep the initial $\textbf{0}$ into a final $\textbf{0}$ in Fig. \ref{figureab1}. The perturbation of the conditions of the bottom particles of the environment around the positions $r_x=8$ nm and $r_x=9$ nm is remarkable. The same happens to the top particles of the environment  around position $r_x=12$ nm. In any case, we see that the environment without reset is able to repeat the operation $\textbf{1}\to \textbf{0}$ correctly for more than 30 times. 

\begin{figure}[ht]
\centering
\includegraphics[width=0.6\linewidth]{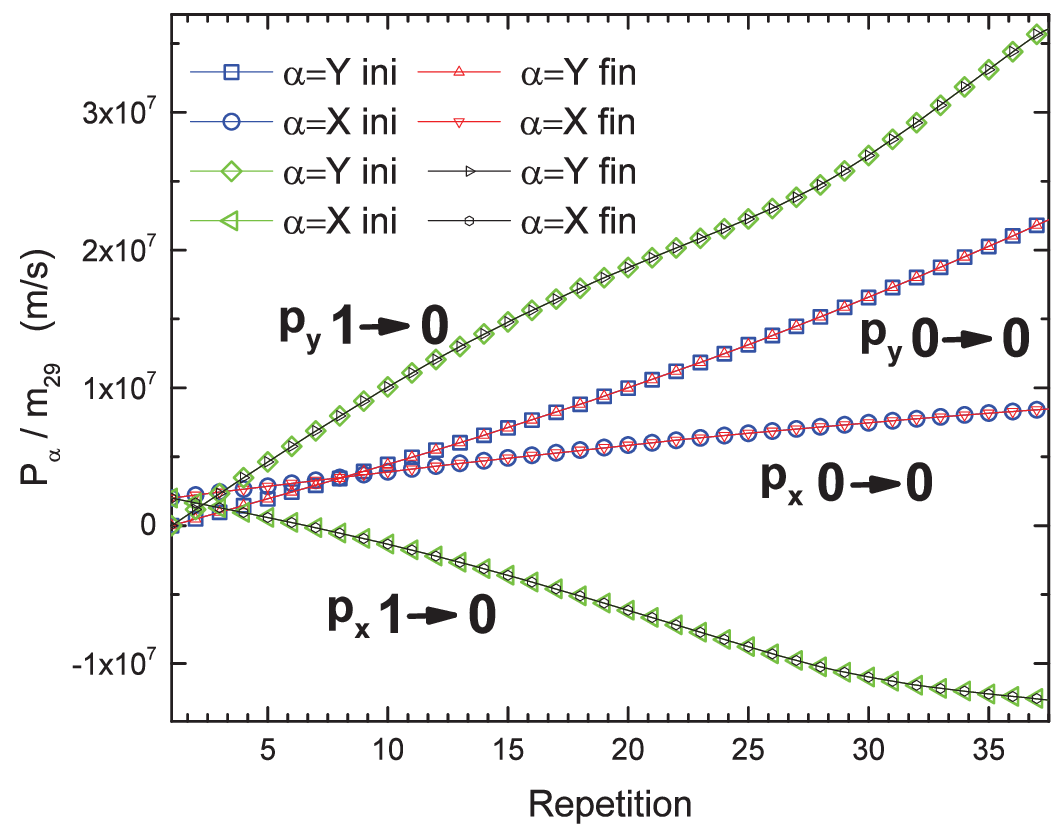}
\caption{Total (initial and final) momentum $p_{\alpha}=\sum_{n=1}^{29} p_{n,\alpha}$ with $\alpha=x$ ($\alpha=y$) for the horizontal (vertical) direction divided by the mass of the system particle (to get units of velocity) for $40$ repetitions of the $\textbf{0}\to \textbf{0}$ operation of Fig \ref{figureab1} or of the $\textbf{1}\to \textbf{0}$ operation of Fig \ref{figureab2}. The final positions and velocities of each particle of the environment after finalizing one operation are the initial conditions of that particle in the next operation. At the beginning of each operation, the system particle has a velocity  $v_{x,29}=2e^6$ m/s and $v_{y,29}=0$. Its initial positions are $r_x=0$ nm and $r_y=3$ nm for defining a $\textbf{0}$ and $r_x=0$ nm and $r_y=7$ nm for defining a $\textbf{1}$. The operation $\textbf{1}\to \textbf{0}$ provokes a greater detectable perturbation of the environment momentum than the operation $\textbf{0}\to \textbf{0}$.}
\label{figureab3}
\end{figure}

In Fig. \ref{figureab3}, we plot the initial and final total momentum of all ($N=29$) particles in the $y$ and $x$ direction for each of the 37 repetitions, for the $\textbf{1}\to \textbf{0}$ and $\textbf{0}\to \textbf{0}$ operations. The first time that an operation takes place, the total momentum in the $y$ direction is zero because none of the particles have velocity in the $y$ direction. The total momentum in the $x$ direction coincides with the initial velocity of the particle of the system (all particles of the environment have zero initial velocity in the $x$ direction). Of course, the final momentum after the operation coincides with the initial one because of the conservation of the total momentum in a closed system. Because of the interactions between particles dictated by \eqref{hamapb}, during the operation, different particles of the environment acquire different velocities. Thus, in the second operation, the system has again the initial velocity $v_{x,29}=2\cdot10^6$ m/s, while the environment particles have the initial conditions at the second operation that correspond to the final conditions of the first operation. A new redistribution of the total momentum happens during the second operation again. All subsequent operations have a similar behavior. 

We clearly see in Fig. \ref{figureab3} that the redistribution of momentum in the $\textbf{1}\to \textbf{0}$ is different than in the $\textbf{0}\to \textbf{0}$ operations. In fact, by just looking at the evolution of the momentum of the environment, without discussing the system dynamics, one can identify which logical operation has occurred in the gate. The environment slightly modifies during  each operation, but the modification of the environment during the operation  $\textbf{0}\to \textbf{0}$ is different from modification of the environment during the operation  $\textbf{1}\to \textbf{0}$. In particular, $p_x$ is negative in $\textbf{1}\to \textbf{0}$ (green triangle and black circle in Fig. \ref{figureab3}), while $p_x$ is positive in $\textbf{0}\to \textbf{0}$ (blue circle and red triangle in Fig. \ref{figureab3}).  

As said in Sec \ref{weak}, these differences between the environments seen in Fig. \ref{figureab3} are exactly what we meant by \textbf{Condition C1} when describing ``environments with different final macroscopic properties''. Of course, such definition can seem a bit ambiguous. Another way of saying the same is that the phase space of the points belonging to the final state of the gate defined by \eqref{hamapb}, when dealing with operations $\textbf{1}\to \textbf{0}$ and $\textbf{0}\to \textbf{0}$, is much more similar to the phase space of Fig. \ref{schemeweak} than to the phase space of Fig. \ref{schemeinter} in the main part of the paper. The later requires a type of chaotic or thermalized behavior of the environment (which can be typical of thermal reservoirs), which is different from the well-defined behavior of the particles of the environment of the toy model that we have worked in this appendix. Again, we conclude with one of the  main messages of this paper: there is no need to assume only chaotic (thermal) environments, since the type of the environments depicted in the phase space of Fig. \ref{schemeweak} is also physically plausible for a logical erasure gate, as seen in Fig. \ref{figureab3}.

\end{appendices}

\bmhead{Supplementary information}

`Not applicable'

\bmhead{Acknowledgments}

X.O. acknowledges  funded by Spain's Ministerio de Ciencia, Innovaci\'on y Universidades under Grant No. RTI2018-097876-B-C21 (MCIU/AEI/FEDER, UE), Grant PID2021-127840NB-I00 (MICINN/AEI/FEDER, UE), the ``Generalitat de Catalunya" and FEDER for the project 001-P-001644 (QUANTUMCAT), the European Union's Horizon 2020 research and innovation programme under Grant No. 881603 GrapheneCore3 and under the Marie Sk\l{}odowska-Curie Grant No. 765426 TeraApps

\section*{Declarations}

\begin{itemize}
\item The authors contributed equally to this work.
\end{itemize}

\section*{Data Availability Statement}

\begin{itemize}
\item Data will be made available on reasonable request.
\end{itemize}


\bibliography{paper}

\begin{thebibliography}{10}
\bibitem{footnote3bis}
When dealing with a thermal environment, it is routinely assumed that the
  entropy variation of the environment is given by $\Delta S_{env}=\Delta Q/T$,
  so that only the entropy variation of the system needs to be evaluated. This
  is the procedure in many classical and quantum developments
  \cite{reeb2014improved}. By contrast, since we are considering general (not
  only thermal) environments, we are discussing the variation of the Boltzmann
  entropy for the whole system plus environment. Our procedure for the direct
  evaluation of the entropy of the whole system has the additional advantage of
  not having to assume that the whole entropy is equal to the sum of the
  entropy of its parts, which is not obvious when the parts have strong
  correlations between them.
\bibitem{footnote5}
Notice that linking a macroscopic property to an expectation value implies that
  such macroscopic property is evaluated over an ensemble of identically
  prepared experiments. In the classical case, macroscopic properties are
  linked to only one experiment. As seen in the appendix, linking a macroscopic
  property to only one experiment is also possible in the quantum regime (after
  properly selecting the quantum interpretation). In any case, the link of a
  macroscopic property to an ensemble-over-experiments allows to provide an
  interpretation-neutral explanation on how the non-universality of Landauer's
  erasure principle is also true in the quantum regime.
\end{thebibliography}


\end{document}